\documentclass[lettersize,onecolumn]{IEEEtran}
\usepackage{amsmath,amsfonts}
\usepackage{algorithmic}
\usepackage{algorithm}
\usepackage{array}
\usepackage{color}
\usepackage[caption=false,font=normalsize,labelfont=sf,textfont=sf]{subfig}
\usepackage{textcomp}
\usepackage{stfloats}
\usepackage{url}
\usepackage{verbatim}
\usepackage{graphicx}
\usepackage{cite}
\usepackage{lmodern}
\usepackage{lineno,hyperref}
\begin{document}

\title{Random walks with resetting on hypergraph}

\author{Fei~Ma,~Xincheng~Hu,~Haobin~Shi,~Wei~Pan,~and~Ping~Wang~\IEEEmembership{Senior Member,~IEEE}
        % <-this % stops a space
\thanks{The research was supported by the National Natural Science Foundation of China No. 62403381, the Fundamental Research Funds for the Central Universities No. G2023KY05105 and the Key Research and Development Plan of Shaanxi Province No. 2024GX-YBXM-021.}% <-this % stops a space
\thanks{Fei Ma is with the School of Computer Science, Northwestern Polytechnical University, Xi'an 710072, China. (e-mail: feima@nwpu.edu.cn).}% <-this % stops a space
\thanks{Xincheng Hu is with the School of Computer Science, Northwestern Polytechnical University, Xi'an 710072, China. (e-mail: xinchenghu@mail.nwpu.edu.cn).}
\thanks{Haobin Shi is with the School of Computer Science, Northwestern Polytechnical University, Xi'an 710072, China. (e-mail: shihaobin@nwpu.edu.cn).}
\thanks{Wei Pan is with the School of Computer Science, Northwestern Polytechnical University, Xi'an 710072, China. (e-mail: panwei@nwpu.edu.cn).}
  \thanks{Ping Wang is with the National Engineering Research Center for Software Engineering, Peking University, Beijing, 100871 China; School of Software and Microelectronics, Peking University, Beijing  102600, China; and Key Laboratory of High Confidence Software Technologies (PKU), Ministry of Education, Beijing, 100871 China. (e-mail: pwang@pku.edu.cn).}
\thanks{Manuscript received xx, 202x; revised xx, 202x.}}

\markboth{Journal of \LaTeX\ Class Files,~Vol.~xx, No.~xx, xx~202x}%
{Shell \MakeLowercase{\textit{et al.}}: A Sample Article Using IEEEtran.cls for IEEE Journals}

\IEEEpubid{0000--0000/00\$00.00~\copyright~202x IEEE}

\maketitle

\begin{abstract}
Hypergraph has been selected as a powerful candidate for characterizing higher-order networks and has received increasing attention in recent years. 
In this article, we study random walks with resetting on hypergraph by utilizing spectral theory. Specifically, we derive exact expressions for some fundamental yet key parameters, including occupation probability, stationary distribution, and mean first passage time, all of which are expressed in terms of the eigenvalues and eigenvectors of the transition matrix. 
Furthermore, we provide a general condition for determining the optimal reset probability and a sufficient condition for its existence.
In addition, we build up a close relationship between random walks with resetting on hypergraph and simple random walks. Concretely, the eigenvalues and eigenvectors of the former can be precisely represented by those of the latter. 
More importantly, when considering random walks, we abandon the traditional approach of converting hypergraph into a graph and propose a research framework that preserves the intrinsic structure of hypergraph itself, which is based on assigning proper weights to neighboring nodes. 
Through extensive experiments, we show that the new framework produces distinct and more reliable results than the traditional approach in node ranking.
Finally, we explore the impact of the resetting mechanism on cover time, providing a potential solution for optimizing search efficiency.
\end{abstract}


\section{Introduction}

%\textcolor[rgb]{1.00,0.00,0.00}{What is the logic in this section? What is the goal of this paper? }

\begin{figure}[h]
    \centering
    \includegraphics[width=\linewidth]{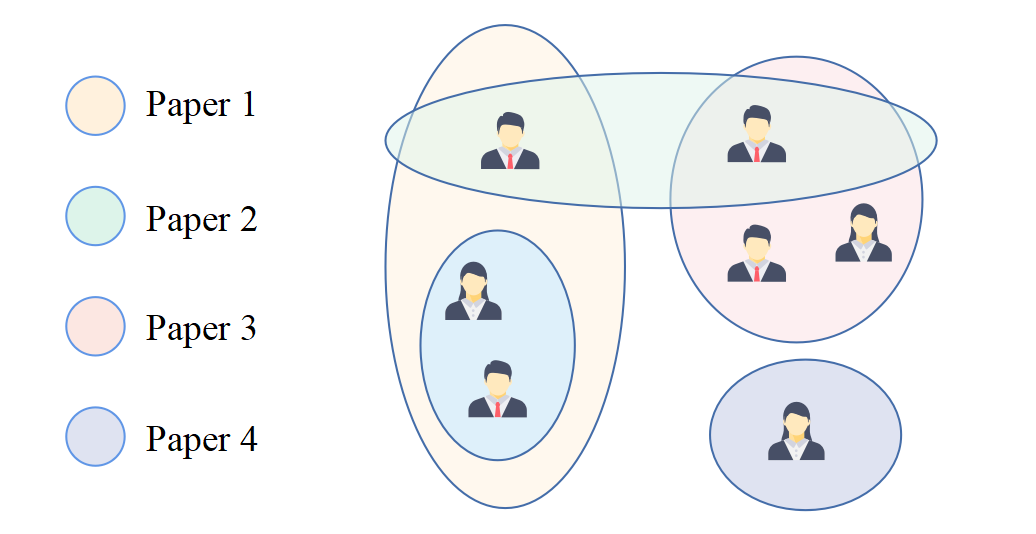}
    \caption{An example hypergraph indicating a co-authorship network where each node represents an author, and each hyperedge connects all the authors who contributed to the same paper.}
    \label{fig:1_1}
\end{figure}

\IEEEPARstart{F}{rom} urban transportation to social systems, and from biology to computer science, networks are ubiquitous in the real world\cite{baumann2020modeling,theodoris2023transfer,avila2020data,ma2024understanding}. Graph is a powerful tool for describing networks, where entities in a network are abstracted as nodes, and the connections between entities are represented as edges. However, graph can only capture the relationships between two entities, failing to express interactions among multiple entities. In reality, many complex networks involve relationships that extend beyond simple pairwise connections\cite{bick2023higher,torres2021and,battiston2020networks}. For instance, in co-authorship networks, multiple authors collaborate on a single paper\cite{huang2020nonuniform}. Complex biochemical reactions often involve more than two substances or reagents\cite{matsuda2020species}. Based on this, hypergraph has been used in order to study these higher-order interactions among multiple nodes. 
In fact, Hypergraph is a generalization of graph, in which (hyper)edges are allowed to connect any number of nodes.\cite{bick2023higher,antelmi2023survey}. A simple hypergraph is illustrated in Fig.\ref{fig:1_1}.

\IEEEpubidadjcol %用来避免版权信息与正文重叠

It is well known that the underlying structure of network significantly influences the dynamical processes they host\cite{castellano2009statistical,arenas2008synchronization,ma2020random}. 
Therefore, some dynamical processes are employed to extract information about the underlying network structure. Among various dynamic processes, random walks\cite{lawler2010random,mastructural} serve as a simple yet powerful tool for uncovering the relational structure of networks. At its core, a random walk describes a path consisting of a succession of random steps within a defined space, where specific transition probabilities determine each move.

Random walks on graph\cite{noh2004random,lovasz1993random} have been extensively studied and employed to calculate node centrality scores\cite{newman2005measure}, identify communities in large-scale networks\cite{rosvall2008maps}, and classify real-world networks\cite{nicosia2014characteristic}. In recent years, as the powerful representational capabilities of hypergraph have gained attention, random walks on hypergraph have gradually become a research hotspot.
For example, a data clustering framework based on random walks on hypergraph has been proposed\cite{hayashi2020hypergraph}. Random walks are utilized to develop the spectral theory of hypergraph with edge-dependent vertex weights\cite{chitra2019random} and to explore the core mechanisms of diffusion processes and biased information propagation in complex networks\cite{carletti2020random}.

Due to the significance and wide applications of random walks in the study of complex network, some variations have been proposed and studied\cite{skardal2019dynamics,ma2022structure,cencetti2018reactive,riascos2020random,tong2008random,evans2011diffusion}. Among which, the approach based on resetting, i.e., interrupted and restarted from a specific node, has been introduced into random walks on graph\cite{riascos2020random,tong2008random,evans2011diffusion}. Interestingly, the average time required to first reach a given target, known as mean first passage time\cite{noh2004random,ma2022random}, can often be minimized by appropriately adjusting the reset rate. In fact, some random search strategies based on resetting have been applied in computer science\cite{luby1993optimal}, statistical physics\cite{montanari2002optimizing} and foraging ecology\cite{boyer2014random}.  However, the reset mechanism has not yet been incorporated into random walks on hypergraph.

%\textcolor[rgb]{1.00,0.00,0.00}{Add additional materials to clarify the importance of random walks with resetting on graph}

In this work, we utilize spectral theory(i.e., eigenvalues and eigenvectors of the transition matrix) to analyze random walks with resetting on hypergraph. We derive general expressions of some key parameters such as occupation probability, stationary distribution, and mean first passage time. 
Furthermore, we study the optimal reset probability that minimizing the mean first passage time and provide general conditions for determining its value, as well as the sufficient condition for its existence, all of which depend only on the first and second moments of the first passage time.
In addition, we prove that random walks with resetting are closely related to simple random walks. Another significant contribution of our work is the proposal of a research framework that preserves the original structure of hypergraph. This is completely different from the traditional approach. The latter is common to convert hypergraph into graph for analysis. Typical conversions include two-section, incidence, and line graph representations\cite{bick2023higher}, though these approaches all disrupt the inherent higher-order interactions.

Besides that it is widely known that random walks can be used to rank nodes based on the stationary distribution\cite{jeh2002simrank,yu2016network}. Through experiments on the DBLP datasets\cite{ley2009dblp}, we prove that our method produces different node ranking results on hypergraph compared with converting hypergraph into two-section graph (i.e., clique graph). This result opens up the possibility for an alternative definition of centrality for higher-order networks including hypergraph. In our second application, we examine the impact of resetting-based approach on network, particularly focusing on how different reset rates affect the cover time\cite{lawler2010random,lovasz1993random,ma2021random}. This investigation offers new insights into optimizing search times in complex networks.

The contribution of this work is as follows.

(1) We employ spectral theory to study random walks on hypergraph in more detail, i.e., using the eigenvalues and eigenvectors of the transition matrix to derive exact expressions of some key parameters, including occupation probability, stationary distribution, and mean first passage time. 

(2) We introduce the concept of random walks with resetting in the study of hypergraph, and also prove that there is a close relationship between this random walk and simple random walks in terms of spectral theory. 

(3) We introduce the concept of the coefficient of variation, which depends solely on the first and second moments of the first passage time. Through this, we establish the general conditions for determining the optimal reset probability, as well as the sufficient condition for its existence.

(4) We propose a research framework that preserves the original structure of hypergraph, avoiding converting it into graph. Specifically, we define a natural and well-suited weighting scheme for neighbour nodes, ensuring that random walks on hypergraph generalize those on graph. 

(5) We conduct extensive experiments on real datasets to demonstrate that the proposed hypergraph research framework produces distinct and reliable node ranking results. Furthermore, we investigate the impact of reset rates on cover time, opening up new avenues for optimizing search times in higher-order networks, including hypergraph.

The rest of this paper is organized as below. 
In Section II, we introduce some definitions and notations, including hypergraph, random walk, and so on. 
In Section III, we study simple random walks on hypergraph in terms of spectral theory and derive occupation probability, stationary distribution, and mean first passage time. 
In Section IV, we incorporate a resetting mechanism into random walk on hypergraph, then build the relationship between transition matrix in this setting and that without resetting. Accordingly, we determine some fundamental parameters.  
In Section V, we study the optimal reset probability and its existence. 
In Section VI, we show some applications based on random walk with resetting to the structural analysis of hypergraph. In Section VII, we review related work. Finally, we close this work in Section VIII.

\section{Definitions and notations}

In this section, we first introduce the definition of hypergraph and a series of matrices related to hypergraph. We then provide some versions of random walk on hypergraph, including simple random walk, generalized random walk and random walk with resetting. 

%analyze why the random walk strategy on hypergraph cannot be simply adapted from the graph random walk strategy, and define a well-suited random walk strategy on hypergraph that considers both the degree of the hyperedges and the number of hyperedges connected to the nodes. Furthermore, when the degree of all hyperedges is 2, this strategy reduces to the graph random walk. Finally, we introduce a resetting mechanism into the random walk.

\subsection{Hypergraph and its matrix representations}

\emph{Graph} is a mathematical model composed of a set of nodes and a set of edges. \emph{Hypergraph} is an extension of graph that relaxes the constraint that an edge can only connect two nodes, allowing edges to connect multiple nodes (these are referred to as hyperedges)\cite{bick2023higher}.

Let us consider hypergraph $\mathcal{H}(V, E)$, where $V = \{1,\ldots,n\}$ is the set of $n$ nodes and $E = \{E_1,\ldots,E_m\}$ is the set of $m$ hyperedges, with $E_{\alpha}$ an unordered collection of nodes.
Therefore a hyperedge is a non-empty subset of the set of nodes, i.e., $E_{\alpha}\subset V$.
When each hyperedge can only connect two nodes, the hypergraph degenerates into a graph.

We can define the \emph{incidence matrix} \cite{carletti2020random} of hypergraph as $\mathbf{e}_{n\times m}$, where each element represents whether a node is connected to a hyperedge, 
\[
    e_{i \alpha}=
    \begin{cases}
        1, & \text{$i \in E_\alpha$} \\
        0, & \text{$i \notin E_\alpha$}
    \end{cases}
\]

With the help of incidence matrix, we define the \emph{adjacency matrix} \cite{carletti2020random} of hypergraph as follows 
\[
    \mathbf{A}_{n \times n}=\mathbf{e}\mathbf{e}^T,
\]
and the \emph{hyperedge matrix} \cite{carletti2020random} of hypergraph
\[
    \mathbf{C}_{m \times m}=\mathbf{e}^T\mathbf{e}.
\]
Thus, $A_{ij}$ represents the number of hyperedges that connect both node $i$ and node $j$.
$C_{\alpha\beta}$ represents the number of nodes in $E_{\alpha}\cap E_{\beta}$.
Specifically, $C_{\alpha\alpha}$ represents the number of nodes in hyperedge $E_\alpha$, which is defined as the degree of $E_\alpha$(i.e., the size of $E_\alpha$).

\subsection{Random walks on hypergraph}

Random walk on hypergraph is one well-studied type of stochastic process that involves moving randomly from one node to an adjacent node of hypergraph.
To describe this process, we need to define transition probability from one node to another node.
A natural and straightforward idea is to define the transition probability as $A_{ij}/k_i$, where $k_i=\sum_{j}A_{ij}$, analogous to the method used for graph. As is well known, this random walk is considered simple, i.e., \emph{simple random walk}\cite{noh2004random,ma2021random-1}. 

However, it is not a wise choice in some cases. This method selects any adjacent node with equal probability without considering the size of the hyperedge (i.e., the degree of the hyperedge) and its impact on the system. It overlooks the higher-order interactions within the same hyperedge. Hence, some variants have been proposed and studied in the literature\cite{skardal2019dynamics,cencetti2018reactive,riascos2020random,tong2008random}. For example, due to group interactions among individuals in social network, the speed of rumor spreading is faster than that of sequential pairwise encounters.
To better characterize the higher-order structure of hypergraph, the walker sitting on a node should assign to all its neighbors a weight that senses the degree of the hyperedges and the number of shared hyperedges. Based on this, we define
\[
    K_{ij}=\sum_{\alpha}(C_{\alpha\alpha}-1)e_{i \alpha}e_{j \alpha},\quad \forall i \neq j.
\]
Note that $K_{ii}=0$. 
$K_{ij}$ encapsulates both the number of hyperedges containing nodes $i$ and $j$, and the number of nodes these hyperedges include, representing a more complex and comprehensive adjacency relationship. 
The matrix $\mathbf{K}$, composed of elements $K_{ij}$, serves as the \emph{generalized adjacency matrix} of hypergraph\cite{carletti2020random}.

Now, we can define transition probability
\[
    W_{ij}=\frac{K_{ij}}{\sum_{l \neq i}K_{il}},
\]
in which term in the denominator is viewed as the hyperdegree of node $i$ in hypergraph and we denote this term by $d_i$, i.e., ${d_i}=\sum_{l\neq i}K_{il}$. 
Thus, the transition probability for hypergraph can be rewritten as
\[
    W_{ij}=\frac{K_{ij}}{d_i}.
\]
The matrix $\mathbf{W}$, composed of elements $W_{ij}$, is the \emph{generalized transition matrix} of hypergraph. As above, the random walk based on matrix $\mathbf{W}$ is referred to as the \emph{generalized random walk} on hypergraph\cite{carletti2020random}. 

It is clear to see that when the degree of all hyperedges is $2$, quantity $K_{ij}$ in hypergraph degenerates into $A_{ij}$ in graph, $d_i$ in hypergraph also degenerates into degree $k_i$ of node in graph.
Thus, it is evident that graph is merely a special case of hypergraph.

As above, we also define other versions of random walk on hypergraph. For example, at each step of random walk, the walker has a probability of $1-\gamma$ to move to one of its neighboring nodes and a probability of $\gamma$ to return to a previously designated node (denoted by $r$ for convenience). In this setting, random walk under consideration is thought of as \emph{random walk with resetting}.  In \cite{riascos2020random}, this random walk on graph has been discussed. Similarly, it is straightforward to see that the corresponding transition matrix $\boldsymbol{\Pi}(r,\gamma)$ is given by 

$$\boldsymbol{\Pi}(r,\gamma)=(1-\gamma)\mathbf{W}+\gamma\boldsymbol{\Theta}(r),$$
where $\mathbf{W}$ is the transition matrix of random walk on hypergraph without resetting as defined above, and $\boldsymbol{\Theta(r)}$ represents a matrix with all elements in the $r$-th column equal to 1 and all other elements equal to $0$. In general, a subset of nodes in hypergraph are chose as candidates for resetting when performing random walk. The associated transition matrix is obtained in a similar manner, which is omitted here.  

Before beginning with our analysis, the primary notations used in this paper are shown in Table \ref{tab:notation}. 
Other notations can be clearly inferred from Refs.\cite{bick2023higher,antelmi2023survey}.

\renewcommand{\arraystretch}{1.5}
\begin{table}[]
    \centering
    \caption{Main notations used in this work}
    \label{tab:notation}
    \begin{tabular}{|c|c|}
    \hline
      Notation & Meanings \\ \hline
    $\mathcal{H}(V,E)$ & Hypergraph\\ \hline
    $\mathbf{e}$ & Incidence matrix of  $\mathcal{H}$ \\ \hline
    $\mathbf{A}$ & Adjacency matrix of  $\mathcal{H}$ \\ \hline
    $\mathbf{C}$ & Hyperedge matrix of  $\mathcal{H}$ \\ \hline
    $\mathbf{K}$ & Generalized adjacency matrix of $\mathcal{H}$ \\ \hline
    $d_i$ & Hyperdegree of node $i$ \\ \hline
    $\mathbf{W}$ & Transition matrix without resetting \\ \hline
    $P_{ij}(t)$ & Occupation probability without resetting\\ \hline
    $\lambda_{l}$ & Eigenvalues of $\mathbf{W}$ \\ \hline
    $\langle \bar{\phi}_l |$ & Left eigenvectors of $\mathbf{W}$ \\ \hline
    $| \phi_l \rangle$ & Right eigenvectors of $\mathbf{W}$ \\ \hline
    $\mathbf{P}$ & Stationary distribution \\ \hline
    $P_j^{\infty}$ & Stationary probability of node $j$\\ \hline
    $\mathcal{N}$ & Sum of the node hyperdegrees \\ \hline
    $\langle T_{ij}\rangle$ & Mean first passage time from $i$ to $j$ \\ \hline
    $r$ & Reset node \\ \hline
    $\gamma$ & Reset rate \\ \hline
    $P_{ij}(t;r,\gamma)$ & Occupation probability with resetting\\ \hline
    $\boldsymbol{\Pi}(r,\gamma)$ & Transition matrix with resetting\\ \hline
    $\zeta_l(r,\gamma)$ & Eigenvalues of $\boldsymbol{\Pi}(r,\gamma)$ \\ \hline
    $\langle \bar{\psi}_l(r,\gamma)|$ & Left eigenvectors of $\boldsymbol{\Pi}(r,\gamma)$ \\ \hline
    $|\psi_l(r,\gamma)\rangle$ & Right eigenvectors of $\boldsymbol{\Pi}(r,\gamma)$ \\ \hline    
\end{tabular}
\end{table}

%The sochastic reset in a random walk is a modification of the traditional random walk model. At each step of the walk, the walker has a probability of $1-\gamma$ to move to a neighboring node and a probability of $\gamma$ to return to the node $r$. This mechanism introduces a "reset" step, effectively enhancing the dynamism and complexity of the random walk.

\section{Some parameters related to simple random walk on hypergraph}

This section aims at studying some fundamental and important parameters pertain to simple random walk on hypergraph. As will be shown in the next section, this is helpful for the study of random walk with resetting on hypergraph.

\subsection{Occupation probability}

In the simple random walk on hypergraph $\mathcal{H}(V, E)$, occupation probability\cite{carletti2020random} as one important parameter represents the probability that after $t$ steps, random walker starting from node $i$ reaches node $j$. This is denoted as $P_{ij}(t)$ in this work.

\textbf{Theorem 1} The occupation probability $P_{ij}(t)$ is given by
\begin{equation}\label{eqa:2024-2-1}
  P_{ij}(t)=\langle i|\phi_1 \rangle \langle \bar{\phi}_1 |j \rangle
             +\sum_{l=2}^{n}\lambda_l^t \langle i| \phi_l \rangle \langle \bar{\phi}_l |j \rangle,
\end{equation}
where $\{\lambda_l\}_{l=2}^n$ are eigenvalues of $\mathbf{W}$, $\langle \bar{\phi}_l |$ and $|\phi_l\rangle$ are, respectively, left eigenvector and right eigenvector corresponding to $\lambda_l$. 
$\langle i|$ is a row vector with the $i$-th element equal to 1 and all other elements equal to 0, while $|j \rangle$ is a column vector with the $j$-th element equal to $1$ and all other elements equal to $0$.

\emph{Proof} By definition, the occupation probability $P_{ij}(t)$ follows the following master equation
\begin{equation}\label{eqa:2024-2-1-1}
    P_{ij}(t+1)=\sum_{l=1}^{n}P_{il}(t)W_{lj},   
\end{equation}
where the transition probability $W_{ij}$ represents the probability of moving from $i$ to $j$ in one step. Using Dirac's notation\cite{griffiths2018introduction}, Eq.(\ref{eqa:2024-2-1-1}) can also be written as
\begin{equation}\label{eqa:2024-2-1-2}
    P_{ij}(t)=\langle i|\mathbf{W}^t|j\rangle.
\end{equation}

Next, we perform spectral decomposition of matrix $\mathbf{W}$, which involves expressing this matrix in terms of its eigenvalues and eigenvectors. In particular, we denote the eigenvalues of $\mathbf{W}$ as $\{\lambda_l\}_{l=1}^n$. Among which, the connectivity of hypergraph $\mathcal{H}(V, E)$ indicates that transition matrix $\mathbf{W}$ has exactly one eigenvalue equal to $1$ while the absolute values of all other eigenvalues are less than $1$. Without loss of generality, we set $\lambda_1=1$. At the same time, it is easy to see that eigenvalue $\lambda_1=1$ has the corresponding right eigenvector proportional to the column vector $(1,1,...,1)^T$.
We generally choose $(1,1,...,1)^T$ and denote it as $|\phi_1\rangle$, and corresponding left eigenvector is denoted as $\langle \bar{\phi}_1 |$. Analogously, for each eigenvalue $\lambda_l$, there is a left eigenvector $\langle \bar{\phi}_l |$ that satisfies $\langle \bar{\phi}_l |\mathbf{W}=\lambda_l \langle \bar{\phi}_l |$,
and a right eigenvector $|\phi_l\rangle$ that satisfies $\mathbf{W}|\phi_l\rangle=\lambda_l|\phi_l\rangle$. For convenience, left and right eigenvectors are orthogonally normalized and also satisfy
\begin{equation}\label{eqa:2024-2-1-3}
  \delta_{ij}=\langle \bar{\phi}_i |\phi_j \rangle,
\end{equation}
and 
\begin{equation}\label{eqa:2024-2-1-4}
    \mathbf{I}=\sum_{l=1}^n|\phi_l\rangle \langle \bar{\phi}_l|,
\end{equation}
where $\delta_{ij}$ is the Kronecker delta function and $\mathbf{I}$ is the identity matrix.
  
Using the notations above, we express matrix $\mathbf{W}$ as follows,

\begin{equation}\label{eqa:2024-2-1-5}
\begin{aligned}
  \mathbf{W}&=\sum_{l=1}^{n}\lambda_l| \phi_l \rangle \langle \bar{\phi}_l | \\
            &=|\phi_1 \rangle \langle \bar{\phi}_1 |+\sum_{l=2}^{n}\lambda_l| \phi_l \rangle \langle \bar{\phi}_l |.
\end{aligned}
\end{equation}

Based on fundamental knowledge of linear algebra, we obtain
\begin{equation}\label{eqa:2024-2-1-6}
\begin{aligned}
  \mathbf{W}^t&=\sum_{l=1}^{n}\lambda_l^t| \phi_l \rangle \langle \bar{\phi}_l | \\
              &=|\phi_1 \rangle \langle \bar{\phi}_1 |+\sum_{l=2}^{n}\lambda_l^t| \phi_l \rangle \langle \bar{\phi}_l |.
\end{aligned}
\end{equation}

Inserting Eq.(\ref{eqa:2024-2-1-6}) into Eq.(\ref{eqa:2024-2-1-2}), we obtain the expression for the occupation probability
\begin{equation}\label{eqa:2024-2-1-7}
\begin{aligned}
  P_{ij}(t)&=\langle i|\mathbf{W}^t|j \rangle \\
           &=\langle i|\phi_1 \rangle \langle \bar{\phi}_1 |j \rangle
             +\sum_{l=2}^{n}\lambda_l^t \langle i| \phi_l \rangle \langle \bar{\phi}_l |j \rangle.
\end{aligned}
\end{equation}
This completes Theorem 1. 

\subsection{Stationary distribution}
The stationary distribution\cite{noh2004random} refers to the distribution of the probabilities of visiting nodes when the random walk reaches a stable state. 
The stationary distribution can be represented by an $n$-dimensional vector $\mathbf{P}=\{P_1^\infty,P_2^\infty,...,P_n^\infty\}$, where $P_j^\infty$ represents the occupation probability as $t$ approaches infinity(i.e., stationary probability), namely, 
\begin{equation}\label{eqa:2024-3-1}
    P_j^\infty=\lim_{t \to \infty} P_{ij}(t).
\end{equation}

\textbf{Theorem 2} The component of the stationary distribution $P_j^{\infty}$ is given by
\begin{equation}\label{eqa:2024-3-2}
    P_j^{\infty}=\frac{d_j}{\mathcal{N}},
\end{equation}
where $\mathcal{N}=\sum_{l=1}^{n}d_l$.

\emph{Proof}
Based on Eq.(\ref{eqa:2024-2-1-1}), we can derive the explicit expression for the occupation probability $P_{ij}(t)$:
\begin{equation}\label{eqa:2024-3-2-1}
    P_{ij}(t)=\sum_{j_1,j_2,\ldots,j_{t-1}}\frac{K_{ij_1}}{d_i} \cdot \frac{K_{j_1j_2}}{d_{j_1}} \cdots \frac{K_{j_{t-1}j}}{d_{j_{t-1}}}.
\end{equation}

Similarly, we derive the explicit expression for $P_{ji}(t)$
\begin{equation}
    P_{ji}(t)=\sum_{i_1,i_2,\ldots,i_{t-1}}\frac{K_{ji_1}}{d_j} \cdot \frac{K_{i_1i_2}}{d_{i_1}} \cdots \frac{K_{i_{t-1}i}}{d_{i_{t-1}}}.
\end{equation}
Here, $j_k$ and $i_k$ are intermediate nodes in the $t$ step transition from $i$ to $j$ or from $j$ to $i$.
Clearly, the intermediate nodes for the transition from $i$ to $j$ are the same as those for the transition from $j$ to $i$.
Therefore, by comparing $P_{ij}(t)$ and $P_{ji}(t)$, we infer that
\begin{equation}\label{eqa:2024-3-2-2}
    d_i P_{ij}(t)=d_j P_{ji}(t).
\end{equation}
Using Eq.(\ref{eqa:2024-3-1}), we obtain $d_i P_j^{\infty}=d_j P_i^{\infty}$, which indicates that in the steady state, the occupation probabilities of nodes are proportional to their hyperdegrees. 
Therefore, by the normalization property of probabilities, we conclude
\begin{equation*}
    P_j^{\infty}=\frac{d_j}{\mathcal{N}},
\end{equation*}
where $\mathcal{N}=\sum_{l=1}^{n}d_l$.

This completes Theorem 2.

In other words, by definition, the stationary distribution $\mathbf{P}$ satisfies the following equation
\begin{equation}
  \mathbf{P}=\mathbf{P}\mathbf{W}.
\end{equation}
Therefore, $\mathbf{P}=\langle\bar{\phi}_1|$. 
Revisiting the first term of the occupation probability $P_{ij}(t)$, whcih is $\langle i|\phi_1 \rangle \langle \bar{\phi}_1 |j \rangle$, reveals that this term equals $P_j^{\infty}$.
Consequently, the occupation probability can also be represented as the probability of the random walker residing at node when the system attains the stationary distribution, namely, 
\begin{equation}\label{eqa:2024-3-3}
  P_{ij}(t)=P_j^{\infty}+\sum_{l=2}^{n}\lambda_l^t \langle i| \phi_l \rangle \langle \bar{\phi}_l |j \rangle.
\end{equation}
\subsection{Mean first passage time}
The mean first passage time \cite{noh2004random} refers to the average number of steps required in a random walk process to travel from a starting node $i$ to first visit another specific node $j$, denoted as $\langle T_{ij} \rangle$.

\textbf{Theorem 3} The mean first passage time $\langle T_{ij} \rangle$ is given by
\begin{equation}\label{eqa:2024-4-1}
    \langle T_{ij} \rangle = 
    \begin{cases}
        \frac{\mathcal{N}}{d_j} & i=j \\
        \frac{\mathcal{N}}{d_j}\left[R_{jj}^{(0)}-R_{ij}^{(0)}\right] & i \neq j 
    \end{cases},
\end{equation}
where $R_{ij}^{(n)}=\sum_{t=0}^{\infty}t^n\{P_{ij}(t)-P_j^{\infty}\}$.

\emph{Proof}
Firstly, we consider the probability $F_{ij}(t)$, that a random walk starting from node $i$ will first reach node $j$ after $t$ steps.
This probability satisfies the following equation
\begin{equation}\label{eqa:2024-4-1-1}
    P_{ij}(t)=\delta_{t0} \delta_{ij}+\sum_{t'=0}^{t} P_{jj}(t-t')F_{ij}(t').
\end{equation}
Then, mean first passage time $\langle T_{ij} \rangle$ from node $i$ to $j$ can be calculated using $F_{ij}(t)$ as below 
\begin{equation}\label{eqa:2024-4-1-2}
    \langle T_{ij} \rangle = \sum_{t=0}^{\infty}t F_{ij}(t).
\end{equation}
In fact, this is the first moment of $F_{ij}(t)$ at the origin.
We introduce the discrete Laplace transform\cite{noh2004random} to facilitate the calculation, i.e., 
\begin{equation}\label{eqa:2024-4-1-3}
    \widetilde{f}(s)=\mathcal{L}\{f(t)\}=\sum_{t=0}^{\infty}e^{-st}f(t).
\end{equation}
By applying Eq.(\ref{eqa:2024-4-1-3}) to Eq.(\ref{eqa:2024-4-1-1}), we obtain
\begin{equation}
\begin{aligned}
     \widetilde{P}_{ij}(s)&=\mathcal{L}\{\delta_{t0}\delta_{ij}\}+\mathcal{L}\{\sum_{t'=0}^{t} P_{jj}(t-t')F_{ij}(t')\} \\
                         &=\delta_{ij}+\widetilde{F}_{ij}(s)\widetilde{P}_{jj}(s).   
\end{aligned}
\end{equation}
This leads to 
\begin{equation}\label{eqa:2024-4-1-4}
    \widetilde{F}_{ij}(s)=(\widetilde{P}_{ij}(s)-\delta_{ij})/\widetilde{P}_{jj}(s).
\end{equation}
We know that $\widetilde{F}_{ij}(s)$ can also be expressed as the Laplace transform of $F_{ij}(t)$ as
\begin{equation}\label{eqa:2024-4-1-5}
    \widetilde{F}_{ij}(s)=\sum_{t=0}^{\infty}e^{-st}F_{ij}(t).
\end{equation}
We next consider the derivative of $\widetilde{F}_{ij}(s)$ with respect to $s$
\begin{equation}
    \frac{d}{ds}\widetilde{F}_{ij}(s)
    =\frac{d}{ds}(\sum_{t=0}^{\infty}e^{-st}F_{ij}(t))
    =\sum_{t=0}^{\infty}(-t)e^{-st}F_{ij}(t).
\end{equation}
By setting $s=0$, we obtain
\begin{equation}
    \left. \frac{d}{ds}\widetilde{F}_{ij}(s)\right|_{s=0}
    =\sum_{t=0}^{\infty}(-t)F_{ij}(t)
    =-\langle T_{ij} \rangle.
\end{equation}
This gives us the formula used to calculate $\widetilde{F}_{ij}(s)$, which is expressed as 
\begin{equation}\label{eqa:2024-4-1-6}
    \langle T_{ij} \rangle = \sum_{t=0}^{\infty}t F_{ij}(t)=-\widetilde{F}_{ij}'(0).
\end{equation}

Obviously, $P_{ij}(t)=P_j^{\infty}+P_{ij}(t)-P_j^{\infty}$. 
For convenience, the first term on the right-hand side is referred to  as the steady-state part, and the sum of the latter two terms as the transient part. We introduce the $n$-th moment of the transient part at the origin
\begin{equation}\label{eqa:2024-4-1-7}
    R_{ij}^{(n)}=\sum_{t=0}^{\infty}t^n\{P_{ij}(t)-P_j^{\infty}\}.
\end{equation}
Next, we calculate the Laplace transform of $P_{ij}(t)$
\begin{equation}\label{eqa:2024-4-1-8}
\begin{aligned}
    \widetilde{P}_{ij}(s) &= \mathcal{L}\{P_{ij}(t)\} \\
                          &= \sum_{t=0}^{\infty}e^{-st}P_j^{\infty}+\sum_{t=0}^{\infty}e^{-st}(P_{ij}(t)-P_j^{\infty}) \\
                          &= P_j^{\infty}\sum_{t=0}^{\infty}e^{-st}+\sum_{t=0}^{\infty}(P_{ij}(t)-P_j^{\infty})\sum_{n=0}^{\infty}\frac{(-s)^n t^n}{n!} \\
                          &= \frac{d_j}{\mathcal{N}(1-e^{-s})}+\sum_{n=0}^{\infty}\frac{(-s)^n}{n!}\sum_{t=0}^{\infty}t^n\{P_{ij}(t)-P_j^{\infty}\} \\
                          &= \frac{d_j}{\mathcal{N}(1-e^{-s})}+\sum_{n=0}^{\infty}(-1)^n R_{ij}^{(n)} \frac{s^n}{n!}
\end{aligned}
\end{equation}

Inserting Eq.(\ref{eqa:2024-4-1-8}) into Eq.(\ref{eqa:2024-4-1-4}) and then taking the derivative together calculate mean first passage time. Two cases need to be discussed in the following. 

In case $i=j$, we have 
\begin{equation}
    \widetilde{F}_{jj}(s)=\frac{\widetilde{P}_{jj}(s)-1}{\widetilde{P}_{jj}(s)}=1-\frac{1}{\widetilde{P}_{jj}(s)},
\end{equation}
and 
\begin{equation}
\begin{aligned}
    \frac{d}{ds}\widetilde{F}_{jj}(s)&=\frac{\widetilde{P}_{jj}'(s)}{\left[\widetilde{P}_{jj}(s)\right]^2}\\
    &=\frac{\frac{d_j}{\mathcal{N}}\cdot\frac{-e^{-s}}{(1-e^{-s})^2}+\sum_{n=1}^{\infty}(-1)^n R_{jj}^{(n)} \frac{s^{n-1}}{(n-1)!}}{\left[\frac{d_j}{\mathcal{N}(1-e^{-s})}+\sum_{n=0}^{\infty}(-1)^n R_{jj}^{(n)} \frac{s^{n}}{(n)!}\right]^2}.
\end{aligned}
\end{equation}
By substituting $s=0$, this yields  
\begin{equation}
\begin{aligned}
    \left. \frac{d}{ds}\widetilde{F}_{jj}(s) \right|_{s=0} &=\lim_{s \to 0}\frac{d}{ds}\widetilde{F}_{jj}(s) \\
    &=\lim_{s \to 0}\frac{\frac{d_j}{\mathcal{N}}\cdot\frac{-e^{-s}}{(1-e^{-s})^2}-R_{jj}^{(1)}s^0}{\left[\frac{d_j}{\mathcal{N}(1-e^{-s})}+R_{jj}^{(0)}s^0\right]^2} \\
    &=-\frac{\mathcal{N}}{d_j}.
\end{aligned}
\end{equation}
Therefore, we obtain 
\begin{equation}
    \langle T_{jj} \rangle = -\widetilde{F}_{jj}'(0) = \frac{\mathcal{N}}{d_j}.
\end{equation}

In case $i\neq j$, a similar analysis leads to 
\begin{equation}
    \widetilde{F}_{ij}=\frac{\widetilde{P}_{ij}(s)}{\widetilde{P}_{jj}(s)},
\end{equation}
and 
\begin{equation}
    \frac{d}{ds}\widetilde{F}_{ij}(s) = \frac{\widetilde{P}_{ij}'(s)\widetilde{P}_{jj}(s)-\widetilde{P}_{ij}(s)\widetilde{P}_{jj}'(s)}{\left[\widetilde{P}_{jj}(s)\right]^2}.
\end{equation}
As above, by performing some simple calculations, we obtain
\begin{equation}
    \langle T_{ij} \rangle = -\widetilde{F}_{ij}'(0) = \frac{\mathcal{N}}{d_j}\left[R_{jj}^{(0)}-R_{ij}^{(0)}\right].
\end{equation}

To sum up, we derive the expression for the mean first passage time as follows 
\[
    \langle T_{ij} \rangle = 
    \begin{cases}
        \frac{\mathcal{N}}{d_j} & i=j \\
        \frac{\mathcal{N}}{d_j}\left[R_{jj}^{(0)}-R_{ij}^{(0)}\right] & i \neq j 
    \end{cases},
\]
When $i=j$, $\langle T_{ij} \rangle$ represents mean first passage time to return to node $j$ for the first time, and hence it is not zero.

This completes Theorem 3.

\section{Random walks with resetting on hypergraph}

Now, we will study random walks with resetting on hypergraph. First, we denote by $P_{ij}(t;r,\gamma)$ the occupation probability of reaching node $j$ after $t$ steps, starting from node $i$, given that the reset node is $r$ and the reset probability is $\gamma$.
It follows the master equation
\begin{equation}\label{eqa:2024-5-0-1}
    P_{ij}(t+1;r,\gamma)=(1-\gamma)\sum_{l=1}^{n}P_{il}(t;r,\gamma)W_{lj}+\gamma\delta_{rj}.
\end{equation}
Accordingly, we have the reset transition matrix $\boldsymbol{\Pi}(r,\gamma)$, whose elements are given by $\Pi_{ij}(r,\gamma)=(1-\gamma)W_{ij}+\gamma\delta_{rj}$.
Thus, the master equation above can be rewritten as
\begin{equation}\label{eqa:2024-5-0-2}
    P_{ij}(t+1;r,\gamma)=\sum_{l=1}^{n}P_{il}(t;,r,\gamma)\Pi_{lj}(r,\gamma).
\end{equation}

As mentioned above, a subset of nodes may be selected as reset candidates. This case is also discussed in the following way. From now on, let us delve into the analysis of random walks with resetting on hypergraph. 

\subsection{Eigenvalues and eigenvectors}
Similarly, we need to analyze the eigenvalues and eigenvectors of the transition matrix $\boldsymbol{\Pi}(r,\gamma)$.
It is straightforward to see that $\boldsymbol{\Pi}(r,\gamma)=(1-\gamma)\mathbf{W}+\gamma\boldsymbol{\Theta}(r)$, where $\mathbf{W}$ is the transition matrix of the random walk on the hypergraph without random reset, and $\boldsymbol{\Theta(r)}$ represents a matrix with all elements in the $r$-th column equal to 1 and all other elements equal to 0.
We denote the eigenvalues of $\boldsymbol{\Pi}(r,\gamma)$ as $\boldsymbol{\zeta}_l(r,\gamma)$, with the corresponding left and right eigenvectors denoted as $\langle \bar{\psi}_l(r,\gamma)|$ and $|\psi_l(r,\gamma) \rangle$ respectively.
Then $\boldsymbol{\Pi}(r,\gamma)$ can be spectrally decomposed in the following form 
\begin{equation}
    \boldsymbol{\Pi}(r,\gamma)=\sum_{l=1}^{n}\boldsymbol{\zeta}_l(r,\gamma)| \psi_l(r,\gamma) \rangle\langle \bar{\psi}_l(r,\gamma)|.    
\end{equation}

In fact, there is a close relationship between the spectral decomposition of $\boldsymbol{\Pi}(r,\gamma)$ and that of $\mathbf{W}$.
For convenience, we first present main results as follows 

\begin{equation}\label{eqa:2024-5-1}
    \zeta_l(r,\gamma)=
    \begin{cases}
        1 & l=1 \\
        (1-\gamma)\lambda_l &l=2,3\ldots,n
    \end{cases},
\end{equation}
\begin{equation}\label{eqa:2024-5-2}
    |\psi_l(r,\gamma)\rangle=
    \begin{cases}
        |\phi_1 \rangle & l=1 \\
        |\phi_l \rangle-\frac{\gamma}{1-(1-\gamma)\lambda_l}\frac{\langle r|\phi_l\rangle}{\langle r|\phi_1\rangle}|\phi_1\rangle
        & l=2,3,\ldots,n
    \end{cases},
\end{equation}
and 
\begin{equation}\label{eqa:2024-5-3}
    \langle \bar{\psi}_l(r,\gamma)|=
    \begin{cases}
        \langle \bar{\phi}_1|+\sum_{m=2}^{n}\frac{\gamma}{1-(1-\gamma)\lambda_m}\frac{\langle r|\phi_m\rangle}{\langle r|\phi_1\rangle}\langle \bar{\phi}_m| &l=1 \\
        \langle \bar{\phi}_l| &l=2,3,\ldots,n
    \end{cases}.
\end{equation}

Concretely speaking, these are shown in the following theorems together with detailed proofs. 

\textbf{Theorem 4} $\boldsymbol{\Pi}(r,\gamma)$ has an eigenvalue $\zeta_1(r,\gamma)=\lambda_1=1$, and the corresponding right eigenvector $|\psi_1(r,\gamma)\rangle=|\phi_1\rangle$.

\emph{Proof}
As mentioned earlier, $|\phi_1\rangle$ is actually $(1,1,...,1)^T$, meaning that all its components are identical. 
Moreover, since $\langle \bar{\phi}_l|\phi_1\rangle=\sum_{i=1}^{n}\langle \bar{\phi}_l|i\rangle\langle i|\phi_1\rangle=\delta_{l1}$, we write
\begin{equation}\label{eqa:2024-5-1-1}
    \sum_{i=1}^{n}\langle\bar{\phi}_l|i\rangle=\frac{\delta_{l1}}{\langle r|\phi_1\rangle}.
\end{equation}
Using Eq.(\ref{eqa:2024-2-1-3}), Eq.(\ref{eqa:2024-2-1-4}), and Eq.(\ref{eqa:2024-5-1-1}) to transform $\boldsymbol{\Theta}(r)$ outputs 
\begin{equation}\label{eqa:2024-5-1-2}
\begin{aligned}
    \boldsymbol{\Theta}(r)&=\mathbf{I} \boldsymbol{\Theta}(r) \mathbf{I} \\
    %&=\sum_{l=1}^{n}|\phi_l \rangle\langle \bar{\phi}_l| \boldsymbol{\Theta}(r) \sum_{m=1}^{n}|\phi_m \rangle\langle \bar{\phi}_m| \\
    &=\sum_{l=1}^{n}\sum_{m=1}^{n} |\phi_l \rangle\langle \bar{\phi}_l| \boldsymbol{\Theta}(r) |\phi_m \rangle\langle \bar{\phi}_m| \\
    %&=\sum_{l=1}^{n}\sum_{m=1}^{n} |\phi_l \rangle\langle \bar{\phi}_l| \sum_{u=1}^{n}|u\rangle\langle u| \boldsymbol{\Theta}(r) \sum_{v=1}^{n}|v\rangle \langle v|\phi_m \rangle\langle \bar{\phi}_m|  \\
    &=\sum_{l=1}^{n}\sum_{m=1}^{n}\sum_{u=1}^{n}\sum_{v=1}^{n} |\phi_l \rangle\langle \bar{\phi}_l|u\rangle\langle u|\boldsymbol{\Theta}(r)|v\rangle \langle v|\phi_m \rangle\langle \bar{\phi}_m| \\
    &=\sum_{l=1}^{n}\sum_{m=1}^{n}\sum_{u=1}^{n}\sum_{v=1}^{n} |\phi_l \rangle\langle \bar{\phi}_l|u\rangle \delta_{vr} \langle v|\phi_m \rangle\langle \bar{\phi}_m| \\
    &=\sum_{l=1}^{n}\sum_{m=1}^{n}\sum_{u=1}^{n} |\phi_l \rangle\langle \bar{\phi}_l|u\rangle \langle r|\phi_m \rangle\langle \bar{\phi}_m| \\
    &=\sum_{l=1}^{n}\sum_{m=1}^{n} |\phi_l \rangle \frac{\delta_{l1}}{\langle r|\phi_1\rangle} \langle r|\phi_m \rangle\langle \bar{\phi}_m| \\
    &=\sum_{m=1}^{n} \frac{\langle r|\phi_m\rangle}{\langle r|\phi_1\rangle}|\phi_1\rangle\langle \bar{\phi}_m|.
\end{aligned}
\end{equation}

Applying Eq.(\ref{eqa:2024-5-1-2}), we obtain
\begin{equation}
\begin{aligned}
    \boldsymbol{\Pi}(r,\gamma)|\phi_1\rangle&=[(1-\gamma)\mathbf{W}+\gamma\boldsymbol{\Theta}(r)]|\phi_1\rangle \\
    &=(1-\gamma)\mathbf{W}|\phi_1\rangle+\gamma\boldsymbol{\Theta}(r)|\phi_1\rangle \\
    &=(1-\gamma)|\phi_1\rangle+\gamma\sum_{m=1}^{n} \frac{\langle r|\phi_m\rangle}{\langle r|\phi_1\rangle}|\phi_1\rangle\langle \bar{\phi}_m|\phi_1\rangle \\
    &=(1-\gamma)|\phi_1\rangle+\gamma\sum_{m=1}^{n} \frac{\langle r|\phi_m\rangle}{\langle r|\phi_1\rangle}|\phi_1\rangle \delta_{m1} \\
    &=(1-\gamma)|\phi_1\rangle+\gamma|\phi_1\rangle \\
    &=|\phi_1\rangle.
\end{aligned}
\end{equation}
This completes Theorem 4.

\textbf{Theorem 5} $\boldsymbol{\Pi}(r,\gamma)$ has eigenvalues $\zeta_l(r,\gamma)=(1-\gamma)\lambda_l$, for $l=2,3,\ldots,n$, and the corresponding left eigenvectors $\langle\bar{\psi_l}(r,\gamma)|=\langle\bar{\phi}_l|$.

\emph{Proof} Applying Eq.(\ref{eqa:2024-5-1-2}), we immediately obtain
\begin{equation}
\begin{aligned}
    \langle \bar{\phi}_l|\boldsymbol{\Theta}(r,\gamma)&= \langle \bar{\phi}_l|[(1-\gamma)\mathbf{W}+\gamma\boldsymbol{\Theta}(r)] \\
    &=\langle\bar{\phi}_l|(1-\gamma)\mathbf{W}+\gamma\langle\bar{\phi}_l|\boldsymbol{\Theta}(r) \\
    &=(1-\gamma)\lambda_l\langle\bar{\phi}_l|+\gamma\sum_{m=1}^{n}\frac{\langle r|\phi_m\rangle}{\langle r|\phi_1\rangle}\langle\bar{\phi}_l|\phi_1\rangle\langle\bar{\phi}_m| \\
    &=(1-\gamma)\lambda_l\langle\bar{\phi}_l|+\gamma\sum_{m=1}^{n}\frac{\langle r|\phi_m\rangle}{\langle r|\phi_1\rangle}\delta_{l1}\langle\bar{\phi}_m| \\
    &=(1-\gamma)\lambda_l\langle\bar{\phi}_l|.
\end{aligned}
\end{equation}

This completes Theorem 5.

\textbf{Theorem 6} The left eigenvector of matrix $\boldsymbol{\Pi}(r,\gamma)$ corresponding to the eigenvalue 1 is given by
\begin{equation}
    \langle\bar{\psi}_1(r,\gamma_1)|=\langle\bar{\phi}_1|+\sum_{m=2}^{n}\frac{\gamma}{1-(1-\gamma)\lambda_m}\frac{\langle r|\phi_m\rangle}{\langle r|\phi_1\rangle}\langle\bar{\phi}_m|.
\end{equation}

\emph{Proof}
Based on Eq.(\ref{eqa:2024-5-1-2}), we have
\begin{equation}
\begin{aligned}
    \langle\bar{\phi}_1|\boldsymbol{\Theta}(r)&=\langle\bar{\phi}_1|\sum_{m=1}^{n}\frac{\langle r|\phi_m\rangle}{\langle r|\phi_1\rangle}|\phi_1\rangle\langle\bar{\phi}_m| \\
    &=\sum_{m=1}^{n}\frac{\langle r|\phi_m\rangle}{\langle r|\phi_1\rangle}\langle\bar{\phi}_m| \\
    &=\langle\bar{\phi}_1|+\sum_{m=2}^{n}\frac{\langle r|\phi_m\rangle}{\langle r|\phi_1\rangle}\langle\bar{\phi}_m|.
\end{aligned}
\end{equation}
When $l=2,3,\ldots,n$, we can further obtain
\begin{equation}
\begin{aligned}
    \langle\bar{\phi}_l|\boldsymbol{\Theta}(r)&=\langle\bar{\phi}_l|\sum_{m=1}^{n}\frac{\langle r|\phi_m\rangle}{\langle r|\phi_1\rangle}|\phi_1\rangle\langle\bar{\phi}_m| \\
    &=\langle\bar{\phi}_l|\phi_1\rangle\sum_{m=1}^{n}\frac{\langle r|\phi_m\rangle}{\langle r|\phi_1\rangle}\langle\bar{\phi}_m| \\
    &=0.
\end{aligned}
\end{equation}
Therefore, we may assume $\langle\bar{\psi}_1(r,\gamma)|=\langle\bar{\phi}_1|+\sum_{m=2}^{n}a_m\langle\bar{\phi}_m|$.
Our goal is to solve for $\{a_m\}_{m=2}^n$.
In addition, it is easy to show that $\langle\bar{\psi}_1(r,\gamma)|\boldsymbol{\Pi}(r,\gamma)=\langle\bar{\psi}_1(r,\gamma)|$, and 
\begin{equation}
\begin{aligned}
    \langle\bar{\psi}_1(r,\gamma)|\boldsymbol{\Pi}(r,\gamma)
    &=\left(\langle\bar{\phi}_1|+\sum_{m=2}^{n}a_m\langle\bar{\phi}_m|\right) \\
    &\quad \times [(1-\gamma)\mathbf{W}+\gamma\boldsymbol{\Theta}(r)] \\
    &=(1-\gamma)\langle\bar{\phi}_1|+\gamma\sum_{m=1}^{n} 
    \frac{\langle r|\phi_m\rangle}{\langle r|\phi_1\rangle}\langle\bar{\phi}_1|\phi_1\rangle\langle \bar{\phi}_m| \\
    &\quad +(1-\gamma)\sum_{m=2}^{n}a_m\lambda_m\langle\bar{\phi}_m| \\
    &=(1-\gamma)\langle\bar{\phi}_1|+\gamma\frac{\langle r|\phi_1\rangle}{\langle r|\phi_1\rangle}\langle\bar{\phi}_1| \\
    &\quad +\gamma\sum_{m=2}^{n} 
    \frac{\langle r|\phi_m\rangle}{\langle r|\phi_1\rangle}\langle\bar{\phi}_1|\phi_1\rangle\langle \bar{\phi}_m| \\
    &\quad +(1-\gamma)\sum_{m=2}^{n}a_m\lambda_m\langle\bar{\phi}_m| \\
    &=\langle\bar{\phi}_1|+\sum_{m=2}^{n}
    \left[\gamma\frac{\langle r|\phi_m\rangle}{\langle r|\phi_1\rangle}+(1-\gamma)a_m\lambda_m\right]\langle\bar{\phi}_m|.
\end{aligned}
\end{equation}
Correspondingly, this requires $a_m=\gamma\frac{\langle r|\phi_m\rangle}{\langle r|\phi_1\rangle}+(1-\gamma)a_m\lambda_m$, which simplifies to $a_m=\frac{\gamma}{1-(1-\gamma)\lambda_m}\frac{\langle r|\phi_m\rangle}{\langle r|\phi_1\rangle}$.
Therefore, we obtain 
\begin{equation*}
    \langle\bar{\psi}_1(r,\gamma)|=\langle\bar{\phi}_1|+\sum_{m=2}^{n}\frac{\gamma}{1-(1-\gamma)\lambda_m}\frac{\langle r|\phi_m\rangle}{\langle r|\phi_1\rangle}\langle\bar{\phi}_m|.
\end{equation*}

This completes Theorem 6.

\textbf{Theorem 7} The right eigenvectors of $\boldsymbol{\Pi}(r,\gamma)$ corresponding to the eigenvalues $(1-\gamma)\lambda_l$, for $l=2,3,\ldots,n$ are given by
\begin{equation}
    |\psi_l(r,\gamma)\rangle=|\phi_l\rangle-\frac{\gamma}{1-(1-\gamma)\lambda_l}\frac{\langle r|\phi_l\rangle}{\langle r|\phi_1\rangle}|\phi_1\rangle.
\end{equation}

\emph{Proof} As above, 
on the basis of Eq(\ref{eqa:2024-5-1-2}), we obtain
\begin{equation}
\begin{aligned}
    \boldsymbol{\Theta}(r)|\phi_l\rangle&=\sum_{m=1}^{n}\frac{\langle r|\phi_m\rangle}{\langle r|\phi_1\rangle}|\phi_1\rangle\langle\bar{\phi}_m|\phi_l\rangle \\
    &=\sum_{m=1}^{n}\frac{\langle r|\phi_m\rangle}{\langle r|\phi_1\rangle}|\phi_1\rangle\delta_{ml} \\
    &=\frac{\langle r|\phi_l\rangle}{\langle r|\phi_1\rangle}|\phi_1\rangle.
\end{aligned}
\end{equation}
This indicates that $|\psi_l(r,\gamma)\rangle=|\phi_l\rangle+b_l|\phi_1\rangle,l=2,3,\ldots,n$.
Our goal is to solve for $\{b_l\}_{l=2}^n$.
It is easy to see that $\boldsymbol{\Pi}(r,\gamma)|\psi_l(r,\gamma)\rangle=(1-\gamma)\lambda_l|\psi_l(r,\gamma)\rangle$, and additionally
\begin{equation}
\begin{aligned}
    \boldsymbol{\Pi}(r,\gamma)|\psi_l(r,\gamma)\rangle
    &=\left[(1-\gamma)\mathbf{W}+\gamma\boldsymbol{\Theta}(r)\right] 
    \left(|\phi_l\rangle+b_l|\phi_1\rangle\right) \\
    &=(1-\gamma)\lambda_l|\phi_l\rangle
    +\gamma\frac{\langle r|\phi_l\rangle}{\langle r|\phi_1\rangle}|\phi_1\rangle \\
    &\quad +(1-\gamma)b_l|\phi_1\rangle
    +\gamma b_l|\phi_1\rangle \\
    &=(1-\gamma)\lambda_l \\
    &\quad \times \left[|\phi_l\rangle+\frac{1}{(1-\gamma)\lambda_l}\left(b_l+\gamma\frac{\langle r|\phi_l\rangle}{\langle r|\phi_1\rangle}\right)|\phi_1\rangle\right].
\end{aligned}
\end{equation}
Correspondingly, this requires $b_l=\frac{1}{(1-\gamma)\lambda_l}\left(b_l+\gamma\frac{\langle r|\phi_l\rangle}{\langle r|\phi_1\rangle}\right)$, which simplifies to $b_l=-\frac{\gamma}{1-(1-\gamma)\lambda_l}\frac{\langle r|\phi_l\rangle}{\langle r|\phi_1\rangle}$.
In a word, we have
\[
    |\psi_l(r,\gamma)\rangle=|\phi_l\rangle-\frac{\gamma}{1-(1-\gamma)\lambda_l}\frac{\langle r|\phi_l\rangle}{\langle r|\phi_1\rangle}|\phi_1\rangle.
\]

This completes Theorem 7.

To sum up, we complete the proof of Eq(\ref{eqa:2024-5-1}), Eq(\ref{eqa:2024-5-2}), Eq(\ref{eqa:2024-5-3}). In the next subsections, we will make use of the obtained results to derive exact solutions of some fundamental parameters. 

\subsection{Some fundamental parameters}

With the eigenvalues and eigenvectors derived in the previous subsection, we derive occupation probability, stationary distribution, and mean first passage time.

Eq.(\ref{eqa:2024-5-0-2}) can be rewritten as
\begin{equation}
    P_{ij}(t;r,\gamma)=\langle i|\boldsymbol{\Pi}(r,\gamma)^t|j\rangle.
\end{equation}
Based on the spectral decomposition of $\boldsymbol{\Pi}(r,\gamma)$, we obtain
\begin{equation}
\begin{aligned}
    P_{ij}(t;r,\gamma)&=\langle i|\psi_1(r,\gamma)\rangle\langle\bar{\psi}_1(r,\gamma)|j\rangle \\
    &\quad +\sum_{l=2}^{n}\left[(1-\gamma)\lambda_l\right]^t\langle i|\psi_l(r,\gamma)\rangle\langle\bar{\psi}_l(r,\gamma)|j\rangle .
\end{aligned}
\end{equation}
The first term in the above equation represents the component of the stationary distribution, $P_j^{\infty}=\langle i|\psi_1(r,\gamma)\rangle\langle\bar{\psi}_1(r,\gamma)|j\rangle$.
Utilizing Eq(\ref{eqa:2024-5-2}) and Eq.(\ref{eqa:2024-5-3}), we obtain
\begin{equation}\label{eqa:2024-5-4}
    P_j^{\infty}(r,\gamma)=\frac{d_j}{\mathcal{N}}+\gamma\sum_{l=2}^{n}\frac{\langle r|\phi_l\rangle\langle\bar{\phi}_l|j\rangle}{1-(1-\gamma)\lambda_l}.
\end{equation}
The occupation probability of node $j$ is then given by
\begin{equation}\label{eqa:2024-5-5}
\begin{aligned}
    P_{ij}(t;r,\gamma)&=P_j^{\infty}(r,\gamma) \\
    &\quad +\sum_{l=2}^{n}\left[(1-\gamma)\lambda_l\right]^t \\
    &\quad \times \left[\langle i|\phi_l\rangle\langle\bar{\phi}_l|j\rangle-\gamma\frac{\langle r|\phi_l\rangle\langle\bar{\phi}_l|j\rangle}{1-(1-\gamma)\lambda_l}\right].
\end{aligned} 
\end{equation}

Since mean first passage time is only related to occupation probability and stationary distribution, based on the previous analysis, we can directly obtain the expression of mean first passage time for the random walk with resetting on hypergraph.
\begin{equation}
\begin{aligned}
    \langle T_{ij}(r,\gamma) \rangle = 
    \begin{cases}
        \frac{1}{P_j^{\infty}(r,\gamma)} & i=j \\
        \frac{1}{P_j^{\infty}(r,\gamma)}\left[R_{jj}^{(0)}(r,\gamma)-R_{ij}^{(0)}(r,\gamma)\right] & i \neq j 
    \end{cases}.
\end{aligned} 
\end{equation}
We can further simplify the expression as
\begin{equation}\label{eqa:2024-5-6}
\begin{aligned}
    \langle T_{ij}(r,\gamma) \rangle &=\frac{1}{P_j^{\infty}(r,\gamma)} \\
    &\quad \times \left[\delta_{ij}+R_{jj}^{(0)}(r,\gamma)-R_{ij}^{(0)}(r,\gamma)\right].
\end{aligned}  
\end{equation}

Because
\begin{equation}\label{eqa:2024-5-7}
\begin{aligned}
    R_{ij}^{(0)}(r,\gamma)&=\sum_{t=0}^{\infty}\left[P_{ij}(t;r,\gamma)-P_j^{\infty}(r,\gamma)\right] \\
    &=\sum_{t=0}^{\infty}\sum_{l=2}^{n}\left[\zeta_l(r,\gamma)\right]^t\langle i|\psi_l(r,\gamma)\rangle\langle\bar{\psi}_l(r,\gamma)|j\rangle \\
    &=\sum_{l=2}^{n}\frac{1}{1-\zeta_l(r,\gamma)}\langle i|\psi_l(r,\gamma)\rangle\langle\bar{\psi}_l(r,\gamma)|j\rangle \\
    &=\sum_{l=2}^{n}\frac{1}{1-\zeta_l(r,\gamma)}\\
    &\quad \times \left[\langle i|\phi_l\rangle-\langle i|\frac{\gamma}{1-(1-\gamma)\lambda_l}\frac{\langle r|\phi_l\rangle}{\langle r|\phi_1\rangle}|\phi_1\rangle\right]\langle\bar{\phi}_l|j\rangle \\
    &=\sum_{l=2}^{n}\frac{1}{1-\zeta_l(r,\gamma)}\\
    &\quad \times\left[\langle i|\phi_l\rangle\langle\bar{\phi}_l|j\rangle-\frac{\gamma}{1-(1-\gamma)\lambda_l}\langle r|\phi_l\rangle\langle\bar{\phi}_l|j\rangle\right],
\end{aligned} 
\end{equation}
we can obtain
\begin{equation}
\begin{aligned}
    R_{jj}^{(0)}(r,\gamma)-R_{ij}^{(0)}(r,\gamma)&=\sum_{l=2}^{n}\frac{1}{1-\zeta_l(r,\gamma)}\\
    &\quad \times \left[\langle j|\phi_l\rangle\langle\bar{\phi}_l|j\rangle-\langle i|\phi_l\rangle\langle\bar{\phi}_l|j\rangle\right] \\
    &=\sum_{l=2}^{n}\frac{1}{1-(1-\gamma)\lambda_l}\\
    &\quad \times \left[\langle j|\phi_l\rangle\langle\bar{\phi}_l|j\rangle-\langle i|\phi_l\rangle\langle\bar{\phi}_l|j\rangle\right].
\end{aligned} 
\end{equation}
Substituting into Eq.(\ref{eqa:2024-5-6}), we obtain the other expression for mean first passage time 
\begin{align}\label{eqa:2024-5-8}
    \langle T_{ij}(r,\gamma) \rangle &=\frac{\delta_{ij}}{P_j^{\infty}(r,\gamma)}+\frac{1}{P_j^{\infty}(r,\gamma)} \notag\\
    &\quad \times \sum_{l=2}^{n}\frac{1}{1-(1-\gamma)\lambda_l}\left[\langle j|\phi_l\rangle\langle\bar{\phi}_l|j\rangle-\langle i|\phi_l\rangle\langle\bar{\phi}_l|j\rangle\right].
\end{align}

\section{Optimal resetting for random walks on hypergraph}
MFPT characterizes the average time required to reach a target node. We now investigate the optimal resetting probability $\gamma^*$ that minimizes the MFPT.
For simplicity, we assume that $r=i$ and $i \neq j$.
Eq.(\ref{eqa:2024-5-8}) can be reformulated as
\begin{equation}\label{eqa:2024-6-1}
\begin{aligned}
    \langle T_{ij}(\gamma) \rangle &\equiv \langle T_{ij}(i; \gamma) \rangle \\
    &= \frac{1}{P_j^\infty(i;\gamma)} 
    \sum_{l=2}^N
    \frac{ \langle j | \phi_l \rangle \langle \bar{\phi}_l | j \rangle 
    - \langle i | \phi_l \rangle \langle \bar{\phi}_l | j \rangle }
    {1 - (1 - \gamma) \lambda_l}.
\end{aligned}
\end{equation}
To simplify the notation, we define
\begin{equation}
\begin{aligned}
    \mathcal{C}_{ij}(\gamma) 
    &\equiv \sum_{l=2}^N \frac{1}{1 - (1 - \gamma)\lambda_l} \bigg[ 
    \langle j | \phi_l \rangle \langle \bar{\phi}_l | j \rangle \\
    &\quad - \langle i | \phi_l \rangle \langle \bar{\phi}_l | j \rangle \bigg],
\end{aligned}
\end{equation}
and 
\begin{equation}
\begin{aligned}
    \mathcal{S}_{ij}(\gamma) 
    &\equiv \sum_{l=2}^N \frac{1}{1 - (1 - \gamma)\lambda_l} 
    \langle i | \phi_l \rangle \langle \bar{\phi}_l | j \rangle.
\end{aligned}
\end{equation}
Utilizing Eq.(\ref{eqa:2024-5-4}), Eq.(\ref{eqa:2024-6-1}) can be compactly written as
\begin{equation}\label{eqa:2024-6-2}
\begin{aligned}
    \langle T_{ij}(\gamma) \rangle 
    = \frac{\mathcal{C}_{ij}(\gamma)}{P_j^\infty + \gamma \mathcal{S}_{ij}(\gamma)}, 
\end{aligned}
\end{equation}
By imposing $\frac{d}{d\gamma}\left\langle T_{ij}(\gamma)\right\rangle \bigg|_{\gamma=\gamma^*} = 0$, we can derive the optimal resetting probability $\gamma^*$, and it satisfies
\begin{equation}\label{eqa:2024-6-3}
\begin{aligned}
    &\mathcal{C}_{ij}'(\gamma^*) \bigg[ P_j^\infty(0) + \gamma^* \mathcal{S}_{ij}(\gamma^*) \bigg] \\
    &- \mathcal{C}_{ij}(\gamma^*) \bigg[ \gamma^* \mathcal{S}_{ij}'(\gamma^*) 
    + \mathcal{S}_{ij}(\gamma^*) \bigg] = 0.
\end{aligned}
\end{equation}
The derivatives $\mathcal{C}_{ij}'(\gamma)$ and $\mathcal{S}_{ij}'(\gamma)$ are given by
\begin{equation}
\begin{aligned}
    \mathcal{C}_{ij}'(\gamma) 
    &= -\sum_{l=2}^N \frac{\lambda_l}{\left[ 1 - (1 - \gamma)\lambda_l \right]^2} \bigg[ 
    \langle j | \phi_l \rangle \langle \bar{\phi}_l | j \rangle \\
    &\quad - \langle i | \phi_l \rangle \langle \bar{\phi}_l | j \rangle \bigg],
\end{aligned}
\end{equation}
\begin{equation}
\begin{aligned}
    \mathcal{S}_{ij}'(\gamma) 
    &= -\sum_{l=2}^N \frac{\lambda_l}{\left[ 1 - (1 - \gamma)\lambda_l \right]^2} 
    \langle i | \phi_l \rangle \langle \bar{\phi}_l | j \rangle.
\end{aligned}
\end{equation}

As seen above, determining the exact optimal resetting probability involves solving a complex equation composed of the eigenvalues and eigenvectors of the transition matrix. 
In the following, we derive a simpler criterion to assess whether resetting improves the random walk. This criterion depends solely on the first two moments of the first passage time distribution,
i.e., $\langle T_{ij}^{2}(\gamma)\rangle$ and $\langle T_{ij}(\gamma)\rangle$.

Analogous to Eq.(\ref{eqa:2024-5-7}), the first moment can be derived as
\begin{equation}
\begin{aligned}
    \mathcal{R}_{i j}^{(1)}(i,\gamma) &=\sum_{t=0}^{\infty} t\left(P_{i j}(t;i,\gamma)-P_{j}^{\infty}\right)\\
    &=\sum_{t=0}^{\infty} \sum_{l=2}^{N} t \zeta_{l}^{t}\left\langle i | \psi_{l}\right\rangle\left\langle\overline{\psi}_{l} | j\right\rangle \\
    &=\sum_{l=2}^{N} \frac{\zeta_{l}}{\left(1-\zeta_{l}\right)^{2}}\left\langle i | \psi_{l}\right\rangle\left\langle\overline{\psi}_{l} | j\right\rangle. 
\end{aligned}
\end{equation}
We can readily obtain
\begin{equation}
\begin{aligned}
    \mathcal{R}_{j j}^{(1)}(i ; \gamma)-\mathcal{R}_{i j}^{(1)}(i ; \gamma) &= \sum_{l=2}^{N} \frac{(1 - \gamma)\lambda_{l} \left[ \langle j | \phi_{l} \rangle \langle \overline{\phi}_{l} | j \rangle - \langle i | \phi_{l} \rangle \langle \overline{\phi}_{l} | j \rangle \right]}{\left(1 - (1 - \gamma)\lambda_{l}\right)^{2}}.
\end{aligned}
\end{equation}
Therefore, $\mathcal{C}_{ij}'(\gamma)$ can be rewritten as
\begin{equation}\label{eqa:2024-6-4}
\begin{aligned}
    \mathcal{C}_{i j}'(\gamma) &= -\frac{1}{1-\gamma} \sum_{l = 2}^{N} \frac{(1 - \gamma)\lambda_{l}}{\left[1-(1 - \gamma)\lambda_{l}\right]^{2}}
    \left[ \langle j | \phi_{l} \rangle \langle \overline{\phi}_{l} | j \rangle - \langle i | \phi_{l} \rangle \langle \overline{\phi}_{l} | j \rangle \right] \\
    &= -\frac{1}{1-\gamma}\left[\mathcal{R}_{j j}^{(1)}(i ; \gamma)-\mathcal{R}_{i j}^{(1)}(i ; \gamma)\right]
\end{aligned}
\end{equation}

According to Eq.(\ref{eqa:2024-6-2}), the optimal $\gamma^*$ satisfies the following equality
\begin{equation}
\begin{aligned}
    \mathcal{C}_{ij}(\gamma^*)=\langle T_{ij}(\gamma^*) \rangle \left[P_j^\infty + \gamma^* \mathcal{S}_{ij}(\gamma^*)\right].
\end{aligned}
\end{equation}
Inserting this into Eq.(\ref{eqa:2024-6-3}), we can obtain
\begin{equation}\label{eqa:2024-6-5}
    \mathcal{C}_{i j}'(\gamma^*) = \langle T_{i j}(\gamma^*) \rangle \left[ \gamma^* \mathcal{S}_{i j}'(\gamma^*) + \mathcal{S}_{i j}(\gamma^*) \right]
\end{equation}
Combining Eq.(\ref{eqa:2024-6-4}) and Eq.(\ref{eqa:2024-6-5}), we can derive
\begin{equation}\label{eqa:2024-6-6}
\begin{aligned}
    &\mathcal{R}_{j j}^{(1)}(i ; \gamma^*) - \mathcal{R}_{i j}^{(1)}(i ; \gamma^*) \\
    & = -(1 - \gamma) \langle T_{i j}(\gamma^*) \rangle \left[ \gamma^* \mathcal{S}_{i j}'(\gamma^*) + \mathcal{S}_{i j}(\gamma^*) \right]
\end{aligned}
\end{equation}

Now, revisiting Eq.(\ref{eqa:2024-4-1-4}) and q.(\ref{eqa:2024-4-1-8}), we can obtain
\begin{equation}
\begin{aligned}
    \widetilde{F}_{ij}(s) &= 1 - s \frac{\mathcal{R}_{jj}^{(0)} - \mathcal{R}_{ij}^{(0)} + \delta_{ij}}{P_{j}^{\infty}} \\
    &\quad\quad + \frac{s^{2}}{2} \Bigg[ \frac{(P_{j}^{\infty} + 2\mathcal{R}_{jj}^{(0)}) \delta_{ij}}{(P_{j}^{\infty})^{2}} \\
    &\quad\quad + \frac{P_{j}^{\infty}(\mathcal{R}_{jj}^{(0)} - \mathcal{R}_{ij}^{(0)} + 2(\mathcal{R}_{jj}^{(1)} - \mathcal{R}_{ij}^{(1)}))}{(P_{j}^{\infty})^{2}} \\
    &\quad\quad + \frac{2\mathcal{R}_{jj}^{(0)}(\mathcal{R}_{jj}^{(0)} - \mathcal{R}_{ij}^{(0)})}{(P_{j}^{\infty})^{2}} \Bigg] + O(s^{3}).
\end{aligned}
\end{equation}
By modifying this expression for the case with resetting, and then taking the first two derivatives with respect to $s$, we can obtain Eq.(\ref{eqa:2024-5-6}) and 
\begin{equation}\label{eqa:2024-6-7}
\begin{aligned}
    \langle T_{i j}^{2}(\gamma) \rangle &= \frac{\mathcal{R}_{j j}^{(0)}(i ; \gamma) - \mathcal{R}_{i j}^{(0)}(i ; \gamma)}{P_{j}^{\infty}(i ; \gamma)} + 2 \frac{\mathcal{R}_{j j}^{(1)}(i ; \gamma) - \mathcal{R}_{i j}^{(1)}(i ; \gamma)}{P_{j}^{\infty}(i ; \gamma)} \\
    &\quad + 2 \frac{\mathcal{R}_{j j}^{(0)}(i ; \gamma) \left[ \mathcal{R}_{j j}^{(0)}(i ; \gamma) - \mathcal{R}_{i j}^{(0)}(i ; \gamma) \right]}{P_{j}^{\infty}(i ; \gamma)^{2}}.
\end{aligned}
\end{equation}
Substituting Eq.(\ref{eqa:2024-5-6}) into the above expression and we can obtain(Note that $i \neq j$)
\begin{equation}
\begin{aligned}
    \langle T^2_{ij}(\gamma) \rangle = & \, \langle T_{ij}(\gamma) \rangle + 2 \frac{R^{(1)}_{jj}(i; \gamma) - R^{(1)}_{ij}(i; \gamma)}{P^\infty_j(i; \gamma)} \\
    & + 2 \frac{R^{(0)}_{jj}(i; \gamma)}{P^\infty_j(i; \gamma)} \langle T_{ij}(\gamma) \rangle.
\end{aligned}
\end{equation}
Utilizing Eq.(\ref{eqa:2024-6-6}), we can derive
\begin{equation}
\begin{aligned}
    \langle T^2_{ij}(\gamma) \rangle = & \, \langle T_{ij}(\gamma) \rangle - 2(1 - \gamma) \frac{\gamma S'_{ij}(\gamma) + S_{ij}(\gamma)}{P^\infty_j(i; \gamma)} \\
    & + 2 \frac{R^{(0)}_{jj}(i; \gamma)}{P^\infty_j(i; \gamma)} \langle T_{ij}(\gamma) \rangle.
\end{aligned}
\end{equation}
Because
\begin{equation}\label{eqa:2024-6-8}
\begin{aligned}
    &(1 - \gamma)\left[\gamma \mathcal{S}_{ij}'(\gamma) + \mathcal{S}_{ij}(\gamma)\right] \\
    &= (1 - \gamma) \sum_{l = 2}^{N} \left\{ -\frac{\gamma \lambda_{l}}{\left[1 - (1 - \gamma)\lambda_{l}\right]^{2}}\right. \\
    &\quad\left.+ \frac{1}{1 - (1 - \gamma)\lambda_{l}} \right\} \langle i | \phi_{l} \rangle \langle \overline{\phi}_{l} | j \rangle \\
    &= (1 - \gamma) \sum_{l = 2}^{N} \frac{1 - \lambda_{l}}{\left[1 - (1 - \gamma)\lambda_{l}\right]^{2}} \langle i | \phi_{l} \rangle \langle \overline{\phi}_{l} | j \rangle.
\end{aligned}
\end{equation}
Based on the relationship between the eigenvectors of the matrices $\mathbf{W}$ and $\boldsymbol{\Pi}(i,\gamma)$, we can obtain
\begin{equation}
\begin{aligned}
    &\langle i | \psi_{l}(i ; \gamma) \rangle \langle \overline{\psi}_{l}(i ; \gamma) | j \rangle \\
    &= \left[ \langle i | \phi_{l} \rangle \langle \overline{\phi}_{l} | j \rangle - \gamma \frac{\langle i | \phi_{l} \rangle \langle \overline{\phi}_{l} | j \rangle}{1 - (1 - \gamma)\lambda_{l}} \right] \\
    &= \frac{(1 - \gamma)(1 - \lambda_{l})}{1 - (1 - \gamma)\lambda_{l}} \langle i | \phi_{l} \rangle \langle \overline{\phi}_{l} | j \rangle \quad \text{for } i \neq j.
\end{aligned}
\end{equation}
Hence, Eq.(\ref{eqa:2024-6-8}) can be rewritten as
\begin{equation}
\begin{aligned}
    & (1 - \gamma)[\gamma \mathcal{S}_{ij}'(\gamma) + \mathcal{S}_{ij}(\gamma)] \\
    &= \sum_{l = 2}^{N} \frac{1}{1 - (1 - \gamma)\lambda_{l}} \langle i | \psi_{l}(i ; \gamma) \rangle \langle \overline{\psi}_{l}(i ; \gamma) | j \rangle \\
    &= \mathcal{R}_{ij}^{(0)}(i ; \gamma).
\end{aligned}
\end{equation}

Therefore, in the optimal case, Eq.(\ref{eqa:2024-6-7}) can be expressed in a more compact form:
\begin{equation}\label{eqa:2024-6-9}
\begin{aligned}
    \langle T_{ij}^{2}(\gamma^{*})\rangle &=\langle T_{ij}(\gamma^{*})\rangle + 2\frac{\mathcal{R}_{jj}^{(0)}(i ; \gamma^{*}) - \mathcal{R}_{ij}^{(0)}(i ; \gamma^{*})}{P_{j}^{\infty}(i ; \gamma^{*})}\langle T_{ij}(\gamma^{*})\rangle \\
    &=\langle T_{ij}(\gamma^{*})\rangle + 2\langle T_{ij}(\gamma^{*})\rangle^{2} \quad \text{for } i \neq j
\end{aligned}
\end{equation}

Here, we introduce the coefficient of variation of the first passage time distribution, which is the ratio of the standard deviation to the mean, and is mathematically defined as:
\begin{equation}
    z_{ij}(\gamma) \equiv \frac{\sqrt{\langle T_{ij}^{2}(\gamma)\rangle - \langle T_{ij}(\gamma)\rangle^{2}}}{\langle T_{ij}(\gamma)\rangle}
\end{equation}
From Eq.(\ref{eqa:2024-6-9}), it is clear that, in the optimal case, the following equality holds:
\begin{equation}\label{eqa:2024-6-10}
    z_{ij}^{2}(\gamma^{*}) = 1 + \frac{1}{\langle T_{ij}(\gamma^{*})\rangle}
\end{equation}

The above expression represents the general condition for determining the optimal resetting probability. 
Additionally, we can derive another condition using a similar method, which will tell us when introducing a small resetting probability can improve the random walk process.
We impose that
\begin{equation}\label{eqa:2024-6-11}
    \left.\frac{d}{d \gamma}\langle T_{ij}(\gamma)\rangle\right|_{\gamma \to 0} < 0.
\end{equation}
Obviously, when $\gamma=1$, the walker remains at the initial node at all times, and MFPT becomes infinite.
Therefore, if Eq.(\ref{eqa:2024-6-11}) holds, then $\langle T_{ij}(\gamma)\rangle$ is non-monotonic, and there exists at least one minimum.
This condition is equivalent to
\begin{equation}
    \left.\frac{d}{d \gamma}\langle T_{ij}(\gamma)\rangle\right|_{\gamma \to 0} = \frac{\mathcal{C}_{ij}'(0) P_{j}^{\infty} - \mathcal{C}_{ij}(0) \mathcal{S}_{ij}(0)}{[P_{j}^{\infty}(0)]^{2}} < 0
\end{equation}
or
\begin{equation}
    \mathcal{C}_{ij}'(0) < \langle T_{ij}(0)\rangle \mathcal{S}_{ij}(0)
\end{equation}
Similar to the previous derivation, we can ultimately obtain the simplified condition:
\begin{equation}\label{eqa:2024-6-12}
    z_{ij}^{2}(0) > 1 + \frac{1}{\langle T_{ij}(0)\rangle} \quad \text{for } i \neq j
\end{equation}

In summary, Eq.(\ref{eqa:2024-6-10}) provides the condition for determining the optimal reset probability.
It is related to the coefficient of variation, which measures the relative fluctuation of MFPT.
Eq.(\ref{eqa:2024-6-12}) provides the sufficient condition for the existence of the optimal reset probability.
It indicates that when the fluctuation of the original process is sufficiently large (i.e., coefficient of variation is significantly greater than $1$), reset improves the search efficiency by reducing the impact of long-duration paths.

\section{Applications}
Here, to make the theoretical results more precise, we will show two applications, i.e., node ranking and optimizing cover time. 

\subsection{Node ranking}

Node ranking holds significant importance in many practical applications, such as identifying key individuals in social networks\cite{girvan2002community,borgatti2005centrality,aral2012identifying}, locating the most influential sources in information dissemination\cite{kempe2003maximizing,kitsak2010identification}, and pinpointing nodes crucial to the overall stability of power networks\cite{buldyrev2010catastrophic}. Since PageRank algorithm\cite{brin1998anatomy}, random walks on the network have been routinely applied to rank nodes based on the probability of a walker visiting them\cite{kleinberg1999authoritative,haveliwala2002topic,ying2018graph}. The more frequently a node is visited, the more important or central it is considered.

In this subsection, we show that higher-order interactions can profoundly impact the ranking, resulting from differing outcomes between random walks on hypergraph and those on the corresponding clique graph.

The dataset we used is derived from the DBLP dataset\cite{ley2009dblp}, which represents a collaboration network of academic papers. Based on this dataset, we construct a hypergraph where papers are represented as hyperedges, and the authors in the same paper are  nodes connected by these hyperedges. We identify the largest connected component of the hypergraph and use it as the hypergraph for our experiments. This resulting hypergraph consists of $1,732$ hyperedges and $6,074$ nodes. Additionally, we convert this hypergraph into its corresponding clique graph for comparative analysis.

We compute the stationary distribution $\mathbf{P}$ of random walks on the hypergraph and the stationary distribution $\mathbf{Q}$ of random walks on the corresponding clique graph.
The stationary probabilities of node $i$ in these two types of random walks are denoted as $P_i^{\infty}$ and $Q_i^{\infty}$, respectively.
We normalize the computed stationary probabilities by their respective maximum values to facilitate comparative visualization.

\begin{figure}[h]
    \centering
    \includegraphics[width=\linewidth]{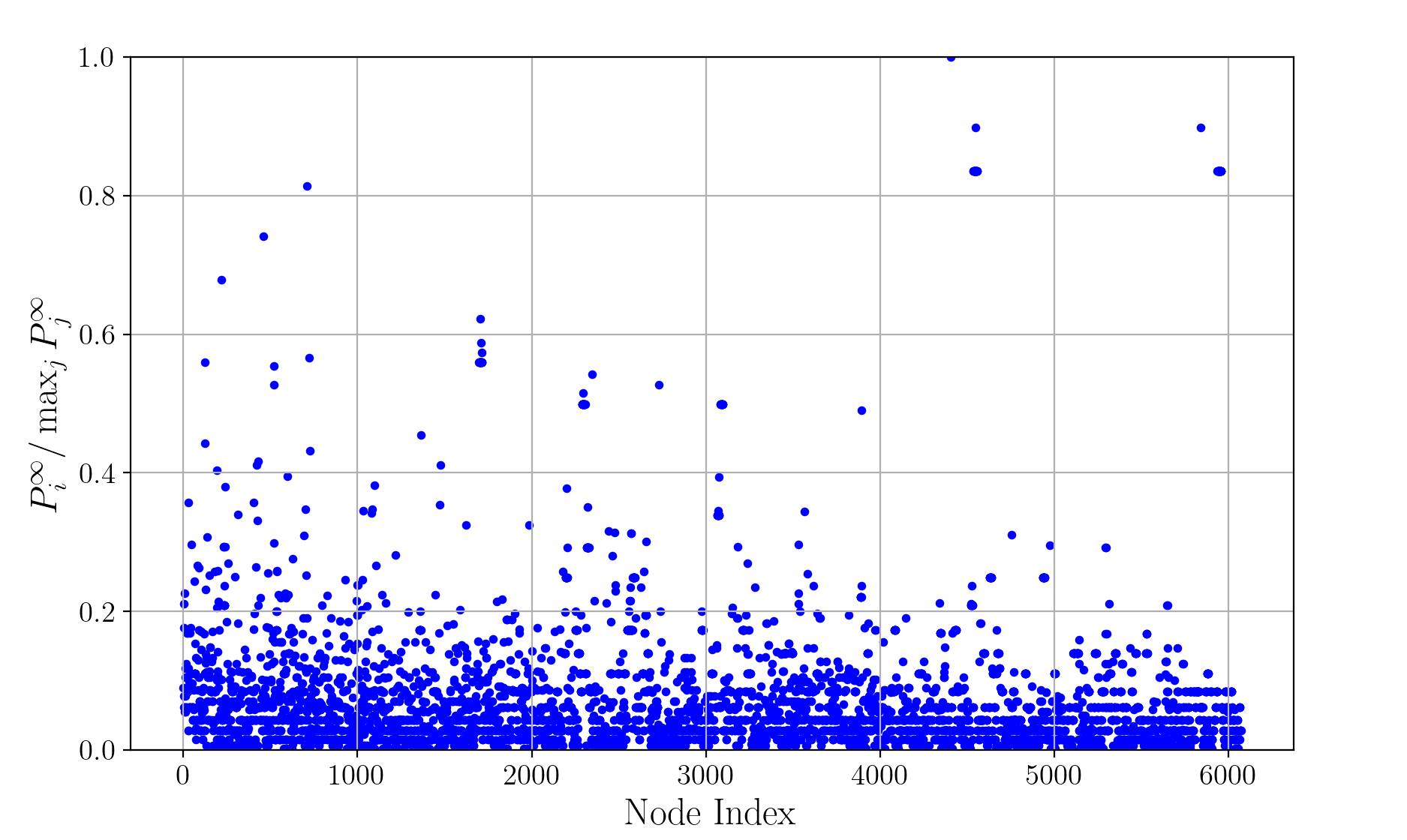}
    \caption{The scatter plot of the normalized ranking obtained from random walks on the hypergraph.}
    \label{fig:5_1}
\end{figure}

\begin{figure}[h]
    \centering
    \includegraphics[width=\linewidth]{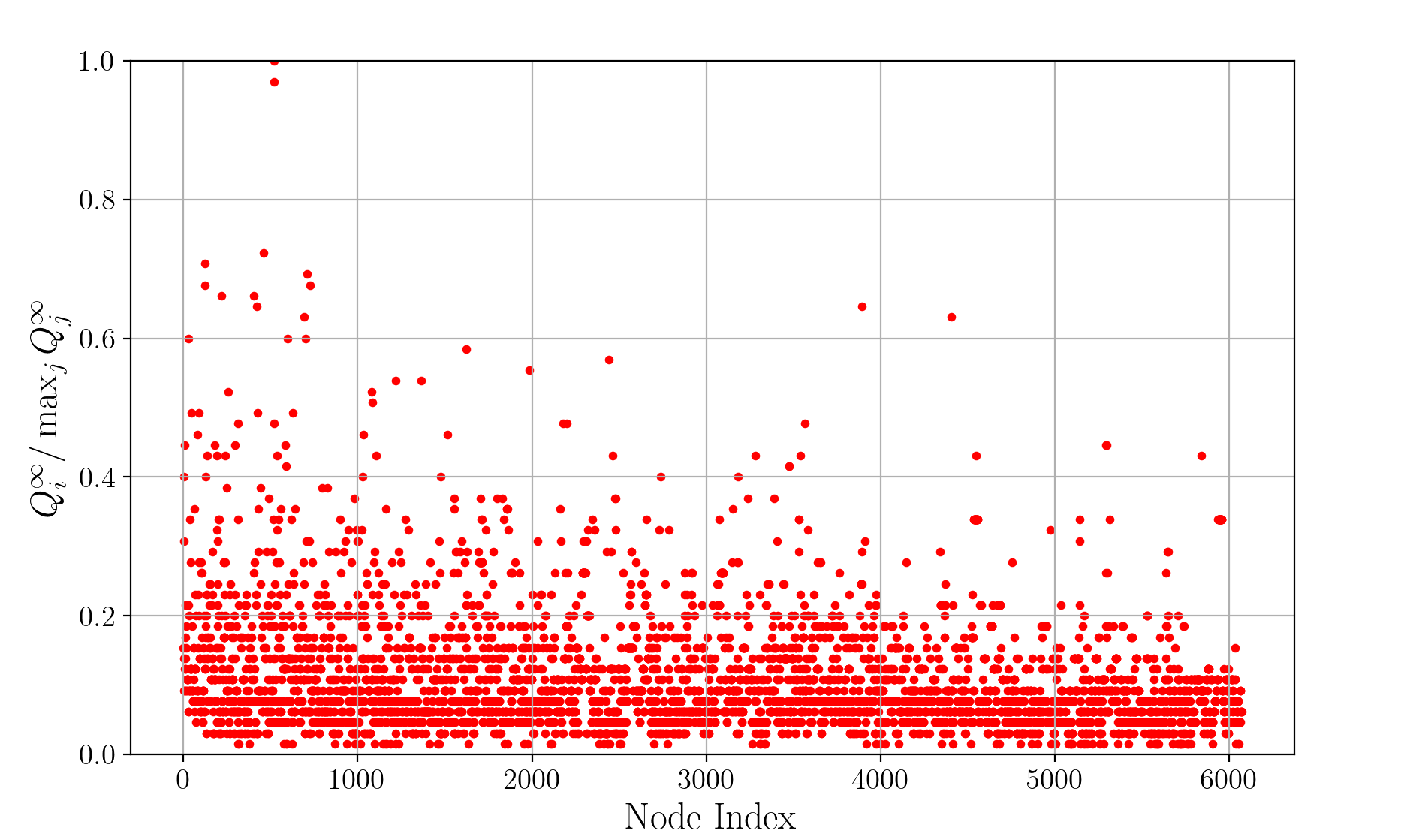}
    \caption{The scatter plot of the normalized ranking obtained from random walks on the corresponding clique graph.}
    \label{fig:5_2}
\end{figure}

Fig.\ref{fig:5_1} shows the relative stationary probabilities of each node in random walk on the hypergraph. Fig.\ref{fig:5_2} displays the relative stationary probabilities of each node in random walk on the corresponding clique graph. From two figures, we can observe that the stationary probabilities obtained from random walks on two structures differ for nodes with the same index. Consequently, the results of node ranking based on these stationary probabilities also vary, with the hypergraph-based ranking being more sensitive to the organization of groups. 

\begin{figure}[h]
    \centering
    \includegraphics[width=\linewidth]{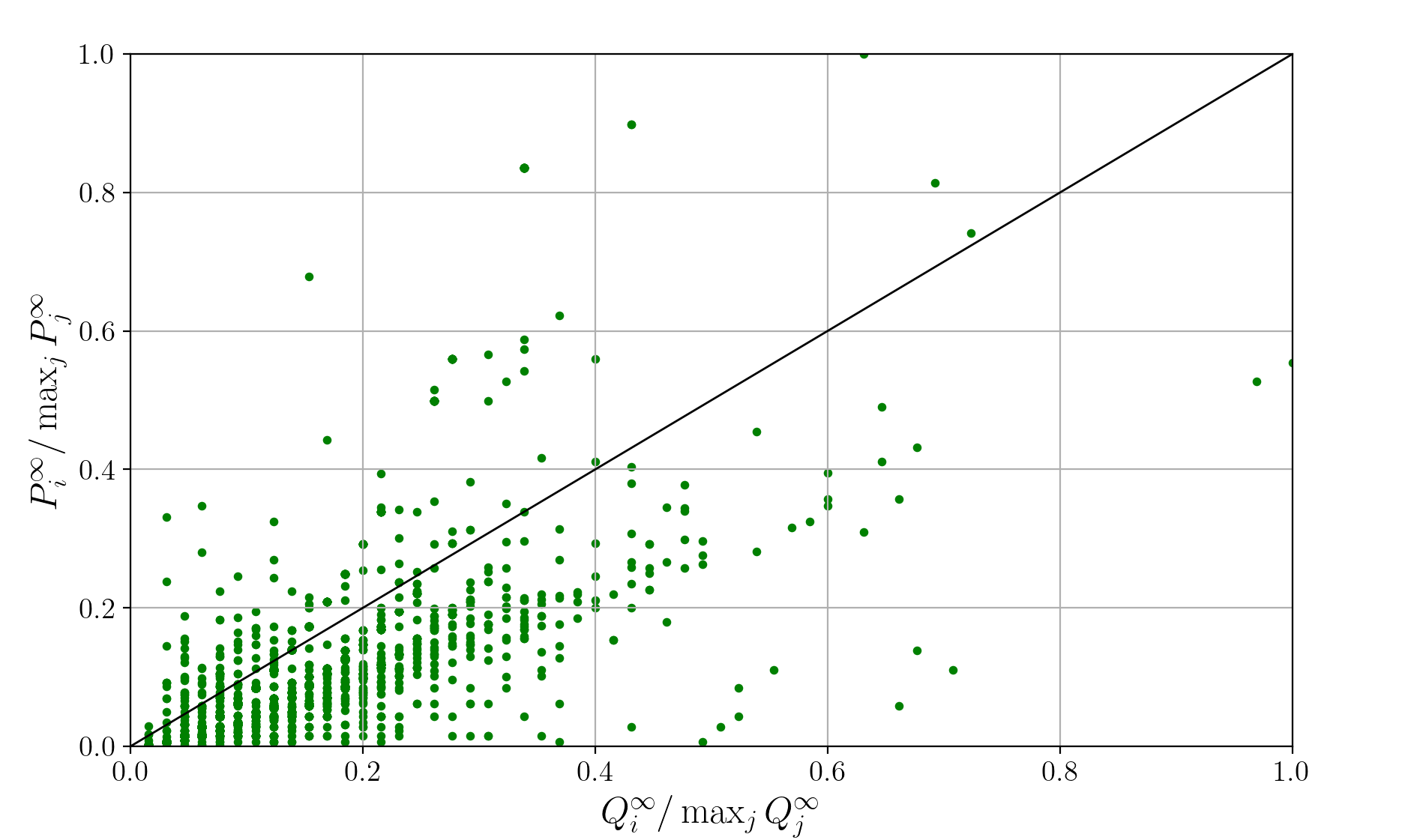}
    \caption{Comparison of the normalized ranking $P_i^{\infty} / \max_j P_j^{\infty}$ on the hypergraph and $Q_i^{\infty} / \max_j Q_j^{\infty}$ on the corresponding clique graph.}
    \label{fig:5_3}
\end{figure}

As a further consideration, we conduct a direct comparison of the results from random walks on two types of representations of hypergraph in question. In Fig.\ref{fig:5_3}, we plot $P_i^{\infty} / \max_j P_j^{\infty}$ vs $Q_i^{\infty} / \max_j Q_j^{\infty}$. If the computed rankings are identical, the data points will lie along the diagonal, deviation from this indicates the new information conveyed by random walks on the hypergraph. Most nodes are located below the diagonal to the right, indicating that their rankings are higher in random walk on the corresponding clique graph. This suggests that these authors have written many papers but with relatively few co-authors. Conversely, nodes located above the diagonal to the left have higher rankings in random walk on the hypergraph, indicating that these authors have participated in a smaller number of papers, but those papers were co-authored by many contributors.  The distribution of nodes on either side of the diagonal reflects a tendency among researchers to prioritize publishing more papers rather than collaborating with a larger number of co-authors. As a consequence, this comparison reflects that it is important to understand nodes in hypergraph using random walks on its own original structure instead of its clique graph. It should be mentioned that a similar analysis may also be used to discuss other representations of hypergraph, which is omitted here due to space limitation. 

\subsection{Optimizing cover time}
In the previous sections, we define mean first passage time (MFPT). Now, we fix the initial node and focus on the average number of steps required for a random walk to visit all nodes, which corresponds to the maximum MFPT for all nodes. This parameter is refer to as the cover time\cite{lawler2010random,lovasz1993random}. 

In computer networks, cover time can be used to measure the efficiency of information dissemination or data collection\cite{chaintreau2007impact,shu2010secure,dembo2004cover}. A shorter cover time for a random walk indicates that information or data can reach all nodes in the network more quickly, thereby enhancing the network's coverage efficiency.

Studies have shown that an appropriate reset rate can optimize random walks on graph, reducing cover time\cite{evans2011optimal,evans2011diffusion,reuveni2016optimal,evans2020stochastic}. In this subsection, we demonstrate that an appropriate reset rate can also optimize the cover time for random walks on hypergraph.

\begin{figure}[h]
    \centering
    \includegraphics[width=\linewidth]{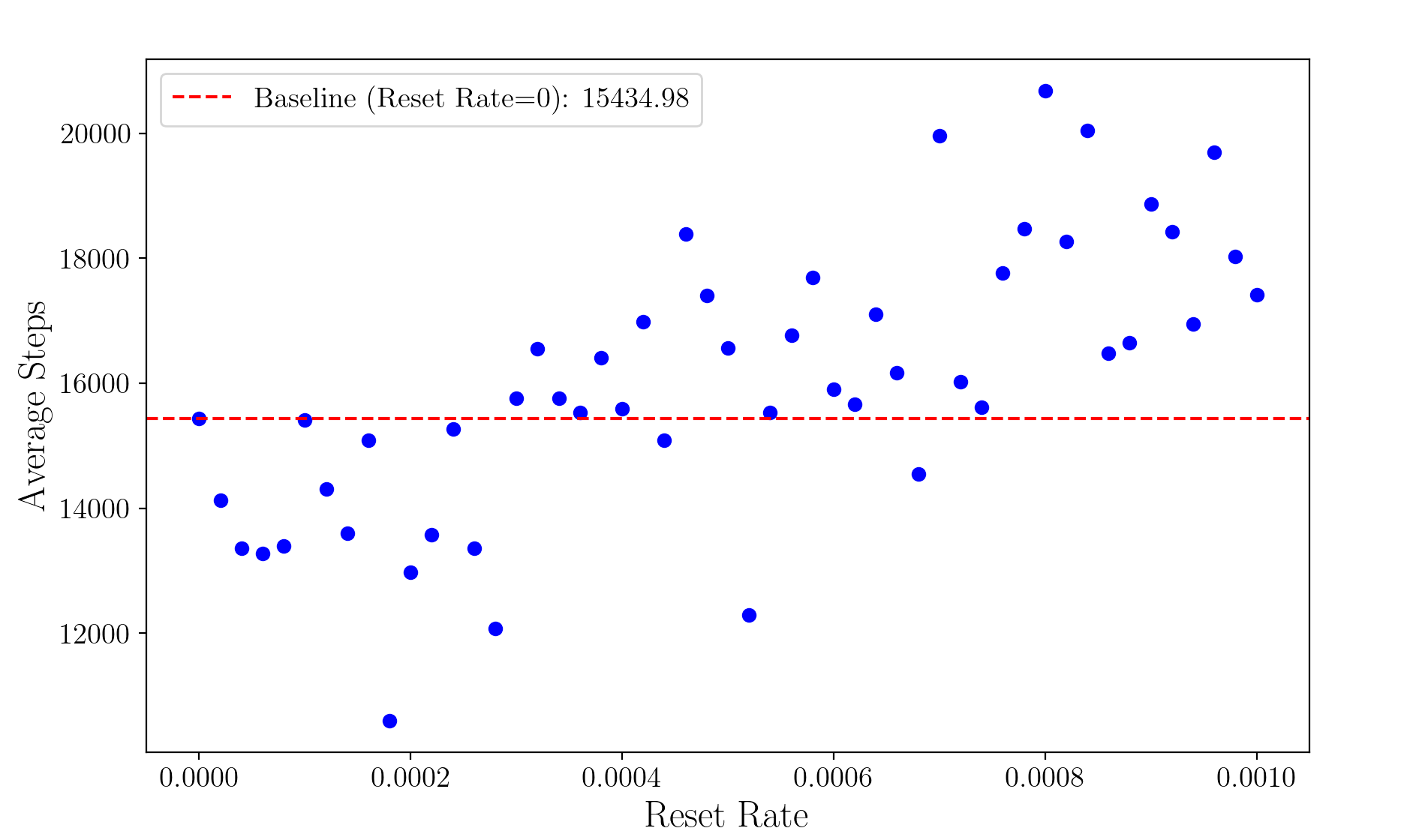}
    \caption{Cover time under different reset rates. The reset rate ranges from $0$ to $0.001$, with a step size of $0.00002$. The red dashed line indicates case $\gamma=0$.}
    \label{fig:5_4}
\end{figure}

To maximize the impact of the reset rate, we fix the initial node (also the reset node) as the node with the highest hyperdegree. The reset rate range is set from 0 to $0.001$, with a step size of $0.00002$. For each reset rate, 50 random walks are conducted, and the average cover time is calculated. Below, we choose to conduct our experiments on a hypergraph consisting of 40 hyperedges and 171 nodes for convenience. The experimental results are shown in Fig.\ref{fig:5_4}. 

Basically, the cover time increases with the rise in reset rate. However, when the reset rate is relatively low, there are certain optimal values where the cover time is shorter than in the case without resetting.

Determining the reset rate that minimizes cover time is a problem of substantial significance. In practice, we carefully search for potential reset rates by setting a small step size, but this approach is not always reliable. Theoretically, the optimal reset rate can be determined by taking the partial derivative of Eq.(\ref{eqa:2024-5-7}). 
Whether the optimal reset rate can be precisely expressed using the eigenvectors and eigenvalues of the transition matrix is an important question that warrants further investigation.

\section{Related work}
In recent years, there has been a surge of interest in complex networks, particularly in higher-order networks that encompass multivariate relationships\cite{bick2023higher,torres2021and,battiston2020networks}. Hypergraph, as a powerful tool for representing higher-order networks, is gaining increasing attention across various fields. Roughly speaking, there are two approaches to studying higher-order networks, including hypergraph. One focuses on the topological structure of network itself\cite{newman2003structure}, while the other examines the dynamic processes that the network hosts\cite{noh2004random,riascos2020random,carletti2020random,ma2024determining}. Our work falls into the latter category.

Random walks, as a simple yet effective dynamic process, have been widely applied in the study of complex networks. Noh et al in\cite{noh2004random} derived an exact expression for mean first passage time between two nodes on graph. In \cite{kahn2000cover}, the authors proposed several exact methods for calculating cover time on graph and discussed the associated bounds. In the past several years, random walks on hypergraph have also garnered research interest\cite{carletti2020random,zhou2006learning}. Traditionally, researchers tend to convert hypergraph into graph for convenience when discussing random walks. Typical conversions include two-section, incidence, and line graph representations\cite{bick2023higher}.
The two-section graph of hypergraph refers to the conversion of each hyperedge into a complete sub-graph, also known as clique expansion\cite{agarwal2006higher,hein2010inverse}.
For the incidence graph of a hypergraph, each hyperedge is converted into a central node, and all the nodes within the hyperedge are connected to this central node. This method transforms a hypergraph into a bipartite graph, also known as star expansion\cite{agarwal2005beyond,karypis1998fast}.
Line graph conversion treats each hyperedge of a hypergraph as a node in a graph, and then creates an edge between two nodes if the corresponding hyperedges share at least one node\cite{yang2022semi,bandyopadhyay2020hypergraph}.
In \cite{agarwal2006higher}, authors performed clique expansion and star expansion on hypergraph, extending the Laplacian operator of random walks to higher-order structures.
However, Hayashi et al in\cite{carletti2020random} proposed a class of random walks directly defined on hypergraph and provided an analytical characterization of this process.
Chitra et al in\cite{chitra2019random} further pointed out that assigning weights to nodes is crucial for random walks on hypergraph, and also emphasized that the weights should depend on the hyperedges, though they did not provide a specific scheme for doing so. In this work, we propose an approach for the assignment of weight in the study of random walks on hypergraph. 

Due to the significance and wide applications of random walks, some variations have been considered\cite{skardal2019dynamics,ma2022structure,cencetti2018reactive,riascos2020random,tong2008random}. Among variations, the resetting mechanism stands out. 
Riascos et al in\cite{riascos2020random} studied random walks with stochastic resetting on graph and applied the results to some certain graphs, such as rings, Cayley trees, and random and complex networks. 
Ref.\cite{tong2008random} designed a fast algorithm based on random walk with resetting to compute the correlation coefficient between two nodes in a weighted graph. 
In \cite{evans2011optimal}, authors explored the impact of different strategies on mean first passage time by setting the reset rate as a function of spatial position.
Evans et al in\cite{evans2011diffusion} proved that resetting has a significant impact on the efficiency of single or multiple searchers in locating a fixed target.
Reuveni \cite{reuveni2016optimal} demonstrated that when a random process is restarted at the optimal reset rate, its relative standard deviation is always equal to $1$.
Ref.\cite{evans2020stochastic} extended the concept of reset to arbitrary stochastic processes (such as Lévy flights or fractional Brownian motion) and non-Poissonian resetting (such as when the time intervals between resetting events follow a power-law distribution). Owing to the inherent complexity of hypergraph, random walks with resetting have not yet been applied to study the higher-order structure of hypergraph.

\section{Conclusion}

We investigate a stochastic process on hypergraph that combines random walk steps to neighbouring nodes and resetting to the initial node.
During exploring the spectral properties of the associated transition matrix, we assume that the probability of the walker’s movement is linearly related to the size of the hyperedge.
Based on this, we propose a research framework that does not require converting hypergraph under consideration into graph.
More specifically, the walker sitting on a node assigns to all its neighbours a weight that senses the size of the hyperedges and the number of shared hyperedges. This strategy turns out to significantly enhances sensitivity to higher-order interactions on hypergraph.

Grounded on our research framework and spectral theory, we derive exact expressions for key parameters such as occupation probability, stationary distribution, and mean first passage time, all of which are explicitly expressed in terms of the eigenvalues and eigenvectors of the transition matrix. 
In addition, we build up a close relationship between random walks with resetting on hypergraph and simple random walks, where the
eigenvalues and eigenvectors of the former can be represented by those of the latter. 
Furthermore, we study the optimal reset probability and its existence. 
Lastly, we also consider two important applications, namely, node ranking and optimizing cover time, which demonstrates the potential behind the theoretical results derived herein.  

%\bibliographystyle{IEEEtran}
%\bibliography{references}

\vfill

\end{document}